\documentstyle[seceq,epsf,psfig,preprint]{ptptex}

\newcommand{\vb}{{\vec b}}
\newcommand{\vp}{{\vec p}}

\newcommand{\vx}{{\vec x}}

\newcommand{\beq}{\begin{equation}}
\newcommand{\eeq}[1]{\label{#1} \end{equation}}

\newcommand{\lton}{\mathrel{\lower.9ex
                  \hbox{$\stackrel{\displaystyle <}{\sim}$}}}
\newcommand{\ee}{\end{equation}} \newcommand{\ben}{\begin{enumerate}}
\newcommand{\een}{\end{enumerate}} \newcommand{\bit}{\begin{itemize}}
\newcommand{\eit}{\end{itemize}} \newcommand{\bc}{\begin{center}}
\newcommand{\ec}{\end{center}} \newcommand{\bea}{\begin{eqnarray}}
\newcommand{\eea}{\end{eqnarray}}
\newcommand{\beqar}{\begin{eqnarray}}
\newcommand{\eeqar}[1]{\label{#1} \end{eqnarray}}

\preprintnumber[3cm]{
CU-TP-974}

\title{
Searching for the Next Yukawa Phase of QCD\footnote{Talk to be published
in the Proc. of 14th Nishinomiya-Yukawa
Memorial Symposium Nov. 1999, Japan; updated with a critique of the CERN press
release Feb. 8, 2000}
}

\author{
Miklos {\sc Gyulassy}}

\inst{
Physics Department, Columbia University,
\\
New York, NY 10027}

\recdate{
\today
}

\abst{
QCD predicts that the 
interactions  between quarks and gluons change from
a confining to a screened Yukawa form
above a critical temperature $T_c\sim
150$ MeV. 
In this talk,
I review some of the key observables in heavy ion reactions 
which are being used
to  search for this  new partonic Yukawa phase at SPS and RHIC. 
These include collective observables such as
$dE_\perp/dyd^2p_\perp$, meson interferometry, jet quenching, and $J/\psi$
suppression.}

\begin{document}

\maketitle

\section{The Ubiquitous Yukawa}
In 1935 H. Yukawa\cite{Yuk35} proposed a theory of nuclear forces based
on the  exchange of a massive boson
\beq
V_Y(r,m)= \alpha_{eff} \frac{e^{-mr}}{r}
\eeq{yukawa}
He estimated that $m \sim 100$ MeV to account 
for the short range $\sim 2$ fm, and in 
1947 Powell  discovered the pion and  confired this theory.
An important theoretical  precursor 
was the  Klein-Gordon
equation \cite{KleinGor}, whose static Green's function
is (\ref{yukawa}).
Elaboration of Yukawa's meson 
theory since then (including spin,
orbit, isospin vertex factors) forms the basis for  the present
effective theory of nuclear forces\cite{Pauli,Bonn}.

In the unrelated field of electrolytes, Debye and H\"uckle
had also  come across the Yukawa potential in another context.
It emerged from solving the
Poisson equation in a conducting medium\cite{Debye}.
Assuming that local charge density fluctuations occur with
a Boltzmann probability, $\exp(-q\phi(x)/T)$,
the  polarizability of the medium in the presence of an external charge
density, leads to a non-linear  self consistent equation
\beq
\nabla \phi(x)=-4\pi (\rho_{ex}(x)+\sum_q q n_q e^{-q\phi(x)/T})
\; \; . \eeq{debye}
In the linearized  approximation, the solution for
a point charge is again eq. (\ref{yukawa}), but in this case
 the effective mass  is the Debye electric screening mass
\beq
\mu^2=4\pi \sum_q q^2 |n_q|/T
\; \; . \eeq{mudeb}
A dense  conductive medium therefore transforms Coulomb into Yukawa.

In nuclear theory, the Yukawa meson mass 
results from the finite gap of the elementary excitations (pions, ...)
of the physical QCD vacuum. By analogy,  it should be possible
to modify the nuclear Yukawa potential by increasing  the nuclear
density or temperature. In this talk I  discuss
current efforts to try to manipulate nuclear matter
in the laboratory to force a breakdown of  Yukawa's hadronic theory.
As we will see, however, it seems difficult  to escape from Yukawa.
In the new deconfined, chirally symmetric phase of QCD at high
temperatures, the Debye-Huckle mechanism takes over and the Yukawa
potential between nucleons mutates into a color-electric screened Yukawa
potential between  partons of a quark-gluon plasma.

\begin{figure}
\epsfxsize= 3 in  
\centerline{\epsfbox{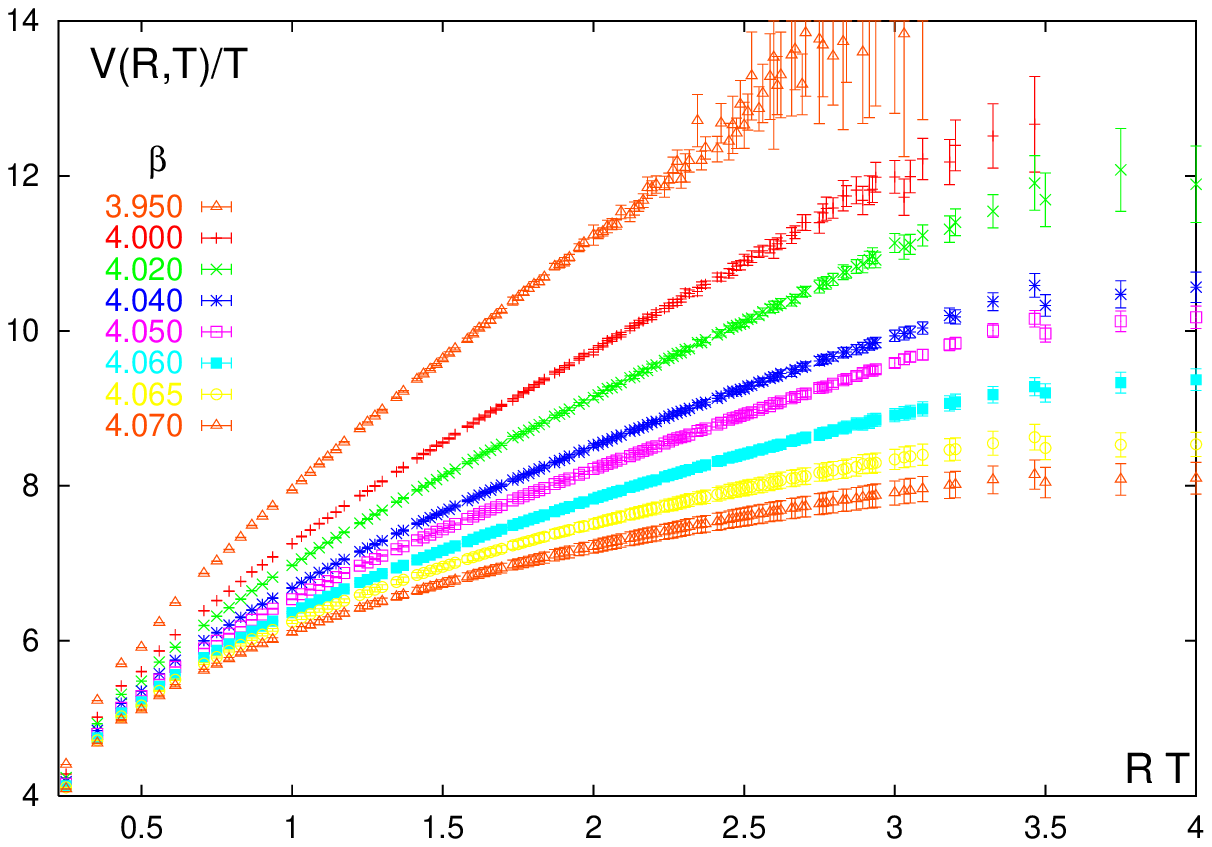}\epsfxsize= 3 in  
\epsfbox{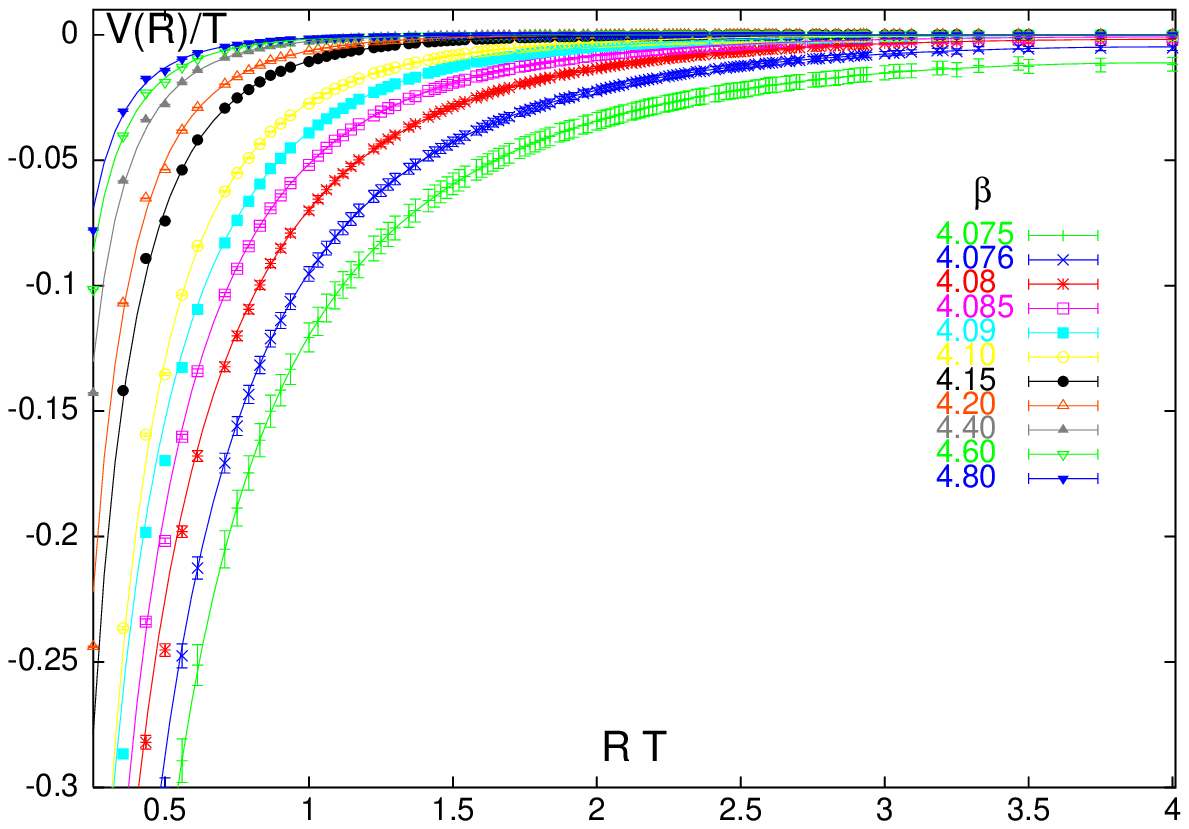}}
\caption{(a) The heavy quark potential in the confined 
phase of SU(2) quenched
QCD (left side)  compared to that in the deconfined Yukawa phase
above \protect{$T_c$} (right side). Results are for a $32^3\times 4$ lattice
from Karsch et al\protect{\cite{Kaczmarek:1999mm}}.
Eq.(\protect{\ref{Gao}}) fits the confined lattice "data" well,
but the QCD string tension decreases more rapidly near $T_c$.
For $T>T_c$ the potential is  screened by the deconfined
gluons (in this quenched calculation) and  acquires
 the generalized Yukawa form
(\protect{\ref{free_d}}). Here $T\approx T_c \exp(11/6(\beta-\beta_c))$ 
with $\beta_c=4.0729$ and the lattice spacing is $a=1/4T$. 
}
\label{fig:confine}
\end{figure}

In QCD, the color potential between partons is approximately
Coulombic at small distances due to the 
asymptotic freedom property of non-Abelian gauge theories.
 However, below a critical temperature, $T_c\sim 150$ MeV,
the confinement property of the {\em nonperturbative} QCD vacuum allows only
 composite color singlet objects (hadrons) to ``roam freely" in the laboratory.
The effective potential between the colored partons  has
a long range linear confining term
$\kappa r$ to prevent them to roaming more than 1 fm away
from any  color neutral blob (hadrons).
The restraining force or "string'' tension, $\kappa\sim 1$ GeV/fm, is huge.
In this confining phase of QCD, the heavy $q\bar{q}$ potential 
is well parameterized by  the
L\"uscher form\cite{Lusch}
\beq
V_{L}(r,0) = -\frac{\alpha_L}{r}+\kappa r
\;\; ,\eeq{lusch}
(as long as dynamical quark pair production is ignored).
The Coulombic part, with strength $\alpha_L=\pi/12$,  arises 
from the zero point quantum fluctuations of the string 
(see p. 803 of Ref\cite{kogut}
 for an  intuitive  derivation). This confining potential 
between heavy quarks has been directly observed
numerically using  lattice QCD techniques\cite{Garden:1999hs}.
As the temperature increases, but remains below
the deconfinement transition, $T<T_c$, the enhanced 
fluctuations due to thermal agitation of the string
modifies the effective potential into the 
approximate Gao form
\cite{Gao}
\begin{eqnarray}
V_G(r,T) & = & -\alpha_L \left[ 1 - \frac{2}{\pi} \tan^{-1}
(2rT)\right]
\frac{1}{r}  +\left[ \kappa - \frac{\pi}{3} T^2(1 - \frac{2}{\pi}
\tan^{-1}(\frac{1}{2rT})\right] r
+ \frac{T}{2} \ln(1+(2rT)^2) \nonumber \\
\label{Gao} 
\end{eqnarray}
The decrease of 
the effective string tension, $\kappa(T)$, predicted above
has been also observed in 
lattice QCD calculations\cite{Kaczmarek:1999mm}  as 
shown in Fig.(\ref{fig:confine}). However, the ``measured''
 string tension is found to
decreases faster than predicted in eq.(\ref{Gao}) near
the  critical temperature. Note that in Fig.(\ref{fig:confine})
$\beta=2N_c/g^2=\beta_c+\frac{12}{11N_c-2N_f}
\log(T/T_c)$, i.e., 
$T\approx T_c \exp(11/6(\beta-\beta_c))$ 
with $\beta_c=4.0729$ for this lattice
calculation.

\begin{figure}
\epsfxsize= 3 in  
\centerline{\epsfbox{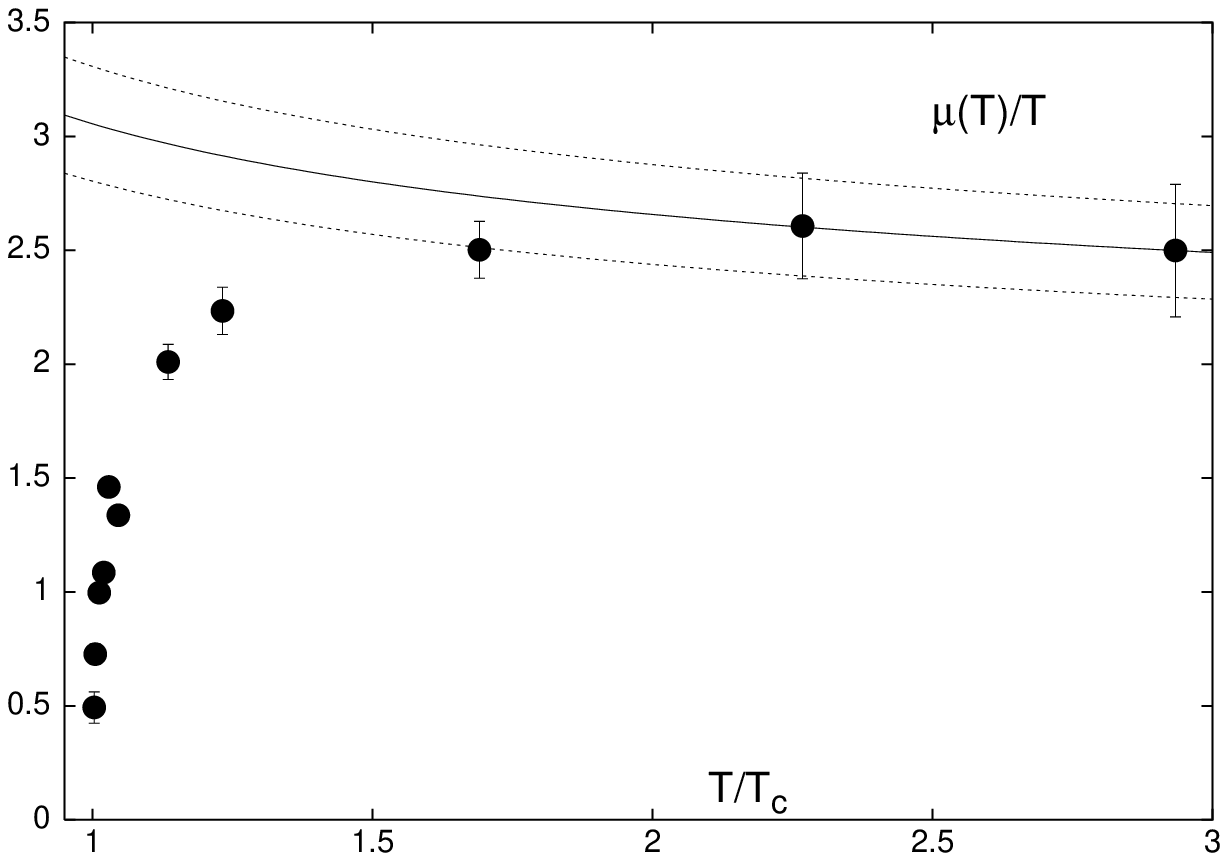}\epsfxsize= 3 in   \epsfbox{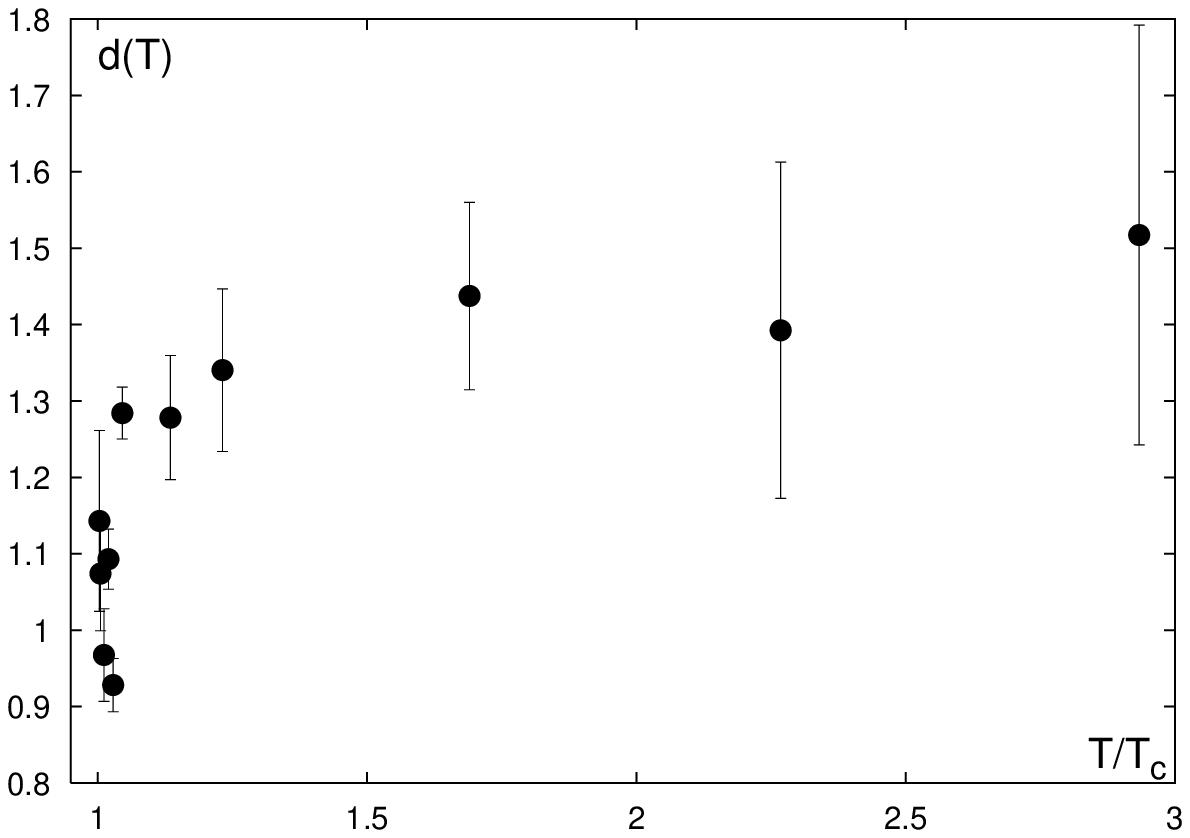}}
\caption{The chromo-electric Debye screening mass, $\mu(T)=2 m_E(T)$   and the
effective exponent, $d(T)$, of the effective Yukawa potential 
in the deconfined  phase versus $T/T_c$ is shown from 
Karsch et al \protect{\cite{Kaczmarek:1999mm}} for SU(2) quenched
QCD. The lines show expected dependence based on thermal  pQCD.
Note that near $T_c$, the range $2/\mu$ is much larger than 
predicted by pQCD and that $d\approx 1$ implies an especially
long range interaction there.
}
\label{fig:mu}
\end{figure}

For temperatures above $T_c$, we see  from Fig.\ref{fig:confine}
that a new  Yukawa phase of QCD is predicted, 
and that the  heavy quark potential mutates into a short range
generalized Yukawa form, which on the lattice is measured in the form 
\beq
V_L(r,T,d) = - \frac{\alpha(T) T}{(rT)^{d_L/2}} e^{-\mu_L(T)r/2}
\;\; .\label{free_d}
\eeq{vy}
Note that $d_L=2, \mu_L(T)=2 m_E(T)$ correspond to a pure Yukawa
interaction in terms of the lattice QCD fit parameters $(d_L,\mu_L)$.
The perturbative
thermal QCD 
 chromo-electric  Debye mass $m_E=\mu(T)/2$ is \cite{Rebhan:1994mx} 
\beq
m_E(T) = g(T)T \left(\frac{N_c}{3}
+\frac{N_F}{6}\right)^{1/2}
\eeq{pertmu}
  for  $N_c$ colors and  $N_F$ flavors. For $N_c=2,N_f=0$ in Fig.
(\ref{fig:confine}), we expect $\mu_L(T)= 2m_E=1.6 g(T) T$
as shown by the solid line in Fig.(\ref{fig:mu}).
The fits to the lattice QCD measurements Fig.(\ref{fig:confine} b)
from \cite{Kaczmarek:1999mm} show that for $T>2 T_c$
 $\mu_L\approx 2.5 T$ is not far from the pQCD estimate.
However, the exponent $d_L\approx 1.5$ is significantly below
the value 2 expected from pQCD. An even more striking nonperturbative
deviation is seen near $T_c$, at which point $d<1$
and $\mu\sim T/2$. This suggests a rather long range interaction
that may be the precursor of the confinement transition.
This strong deviation from the perturbative QCD 
near $T_c$ 
was also reported previously in ref.\cite{Gao:1990br},
where in addition it was found that the effective coupling
is rather small $\alpha\sim 0.15$ above $T_c$. 
It is important to
keep in mind that the above numerical experiments do not include
dynamical quarks and are limited to SU(2). Nevertheless, they provide
strong evidence that QCD
predicts a qualitatively new (nonperturbative) 
partonic Yukawa phase of matter that
should exist at an energy density only an order of
magnitude above that in ground state nuclei ($\epsilon > 2$ GeV/fm$^3$).

\begin{figure}
\epsfxsize= 3. in  
\centerline{\hspace{0.15in}\epsfbox{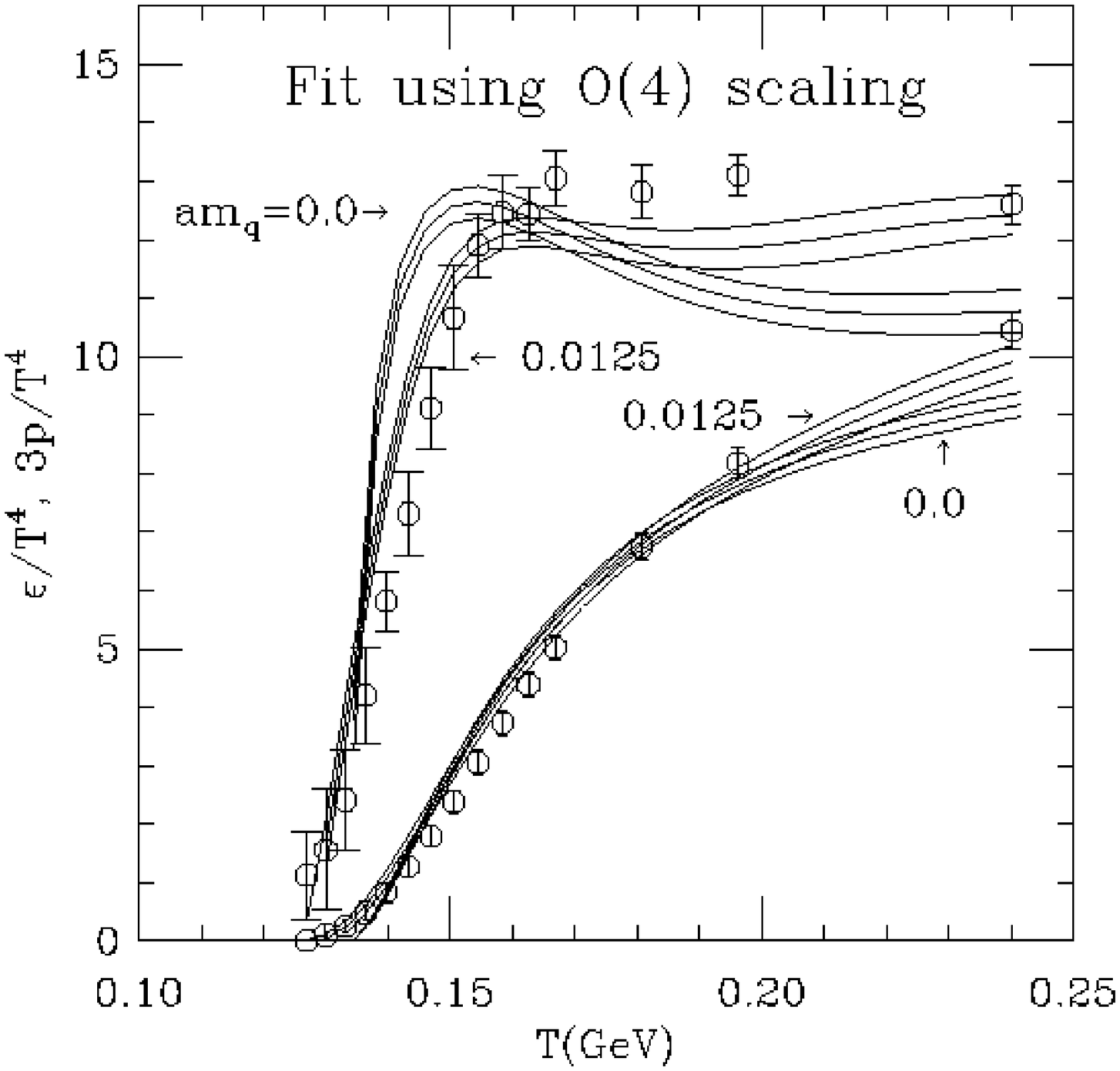}\epsfxsize= 3. in   
\hspace{-0.15in} \epsfbox{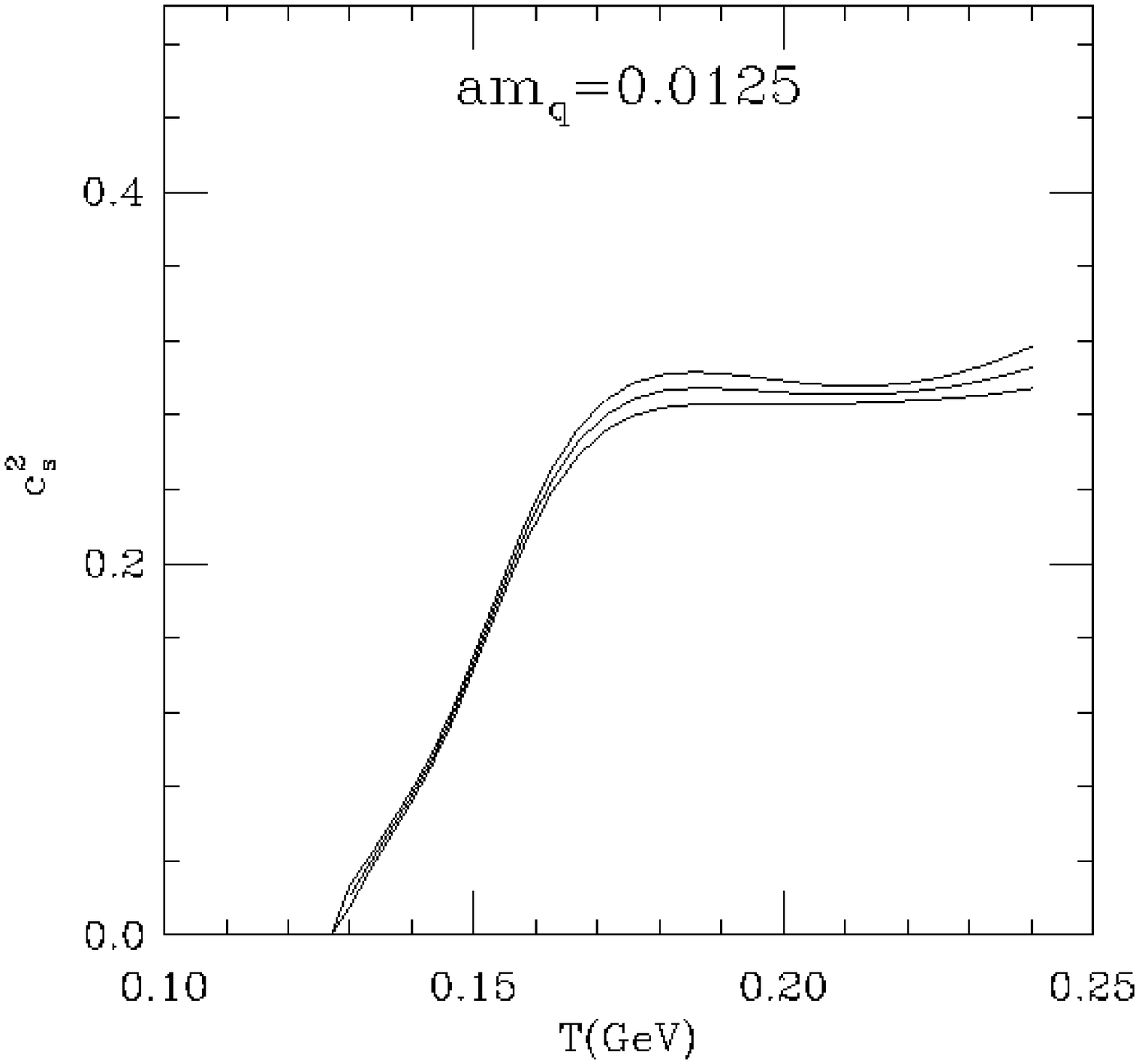}}
\caption{Thermodynamic 
energy density ($\epsilon/T^4$ top curves left), pressure
($3p/T^4$ lower curves left) and speed of sound squared
(right) from lattice QCD
(2 flavor $12^3\times 6$) from the MILC collab \protect{\cite{Bernard:1997cs}}.
The curves are zero quark mass extrapolations.
Note the rapid reduction of the pressure and speed of sound
as $T_c$ is approached from above.}
\label{fig:eos}
\end{figure}

The thermodynamic properties of the deconfined QCD phase
are shown in Fig.\ref{fig:eos} from ref.\cite{Bernard:1997cs}
for 2 flavor $12^3\times 6$ lQCD.
A present limitation of all lattice results so far 
is that the pion is still too massive to make make contact
with the ``known'' thermodynamic
 properties of ordinary hadronic/nuclear matter below $T_c$.
Recent advances in implementing Domain
Wall Fermions on the lattice and the availability 
of new TeraFlop scale computers at Columbia and 
the Riken Brookhaven Research Center\cite{QCDSP}
and the CP-PACS project in Tsukuba\cite{CP-PACS}, 
should enable much more precise 
calculations of the quark-gluon plasma equation of state
in the near future.
Two striking features of Fig.(\ref{fig:eos}) suggest two key
observable signatures of this phase transition in nuclear collisions.
First, the entropy density $\sigma(T)=(\epsilon+p)/T$ increase
very rapidly with $T$ in a narrow interval $\Delta T/T_c< 0.1$.
Second, the plasma becomes extremely soft $p/\epsilon\ll 1$ and
$c_s^2\ll1$ near $T_c$. As we review below, the first feature can lead to
time delay (the QGP stall) 
measurable via hadronic interferometry. The second
feature can lead to interesting non-linear collective 
flow observables in nuclear collisions.
The experimental verification of these fundamental 
predictions of QCD is the primary motivatiion
for the heavy ion experimental program at Brookhaven and CERN.
In the following sections, I review first  how deep into the QGP
phase RHIC may be able to reach,
and then discuss several signatures used to test
the QCD  predictions in such experiments.

\section{Initial conditions in A+A}

In order to see the partonic Yukawa phase, we must first
create an extended blob of matter at 100 times the density of nuclear matter.
Head on collisions of heavy ions are used for that.
There are many ways to describe
how this dense matter is formed.
 One intuitive picture is given in terms of 
the McLerran-Venugopalan model\cite{McLerran:1994ni}.
For highly boosted nuclei with  $E_{cm}\sim 100
m_N$, time dilation effectively freezes out the  quantum chromo fluctuations 
inside the nuclei while the two pass through each other. 
Heavy Au beams can then be regarded as  
 well collimated, ultra dense
beams of partons. This  (chromo Weizsacker-Williams)  gluon 
cloud contains a very large number, $G_A(x,p_0^2)\sim A/x^{1+\delta}$,
of almost on-shell collinear gluons with longitudinal momentum fraction
$x=p_0/E_{cm}\ll 1$. As the clouds pass through each other,
partons scatter via chromo Rutherford and decohere into a mostly 
gluon plasma on a fast time scale $1/p_\perp \ll 1$ fm/c.
  The number of gluons pairs (mini-jets) extracted from the nuclei
by this mechanism at
 rapidities $y_i$ and transverse momentum $\pm k_\perp$ 
can be calculated in pQCD as follows
\cite{Eskola:2000fc,Eskola:1989yh,Blaizot:1987nc,Gyulassy:1997vt}
\beq
\frac{dN_{AB\rightarrow ggX}}{dy_1dy_2dk_\perp^2}
=K x_1 G_A(x_1,k_\perp^2) x_2 G_B(x_2,k_\perp^2)
\frac{d\sigma_{gg\rightarrow gg}}{dk_\perp^2}
T_{AB}(\vb)
\;\; ,\eeq{mini}
where $x_1=x_\perp(\exp(y_1)+\exp(y_2))$ and
$x_2=x_\perp(\exp(-y_1)+\exp(-y_2))$, with $x_\perp=k_\perp/\surd s$, and where
the pQCD $gg\rightarrow gg $ cross section for scattering with
$t=-k_\perp^2(1+\exp(y_2-y_1))$ and $y_2-y_1=y$ is given by \beqar
\frac{d\sigma^{gg}}{dt}&=& 
\frac{9}{8}
\frac{4\pi\alpha^2}{k_\perp^4}\frac{(1+e^y+e^{-y})^3}{(e^{y/2}+e^{-y/2})^6}
\; \; . \eeqar{ggqcd}
The nuclear baryon number, $A$, only plays here the role
of increasing the density of partons and providing the geometrical
amplification, $T_{AB}(\vb)\stackrel{<}{\sim} 30/$mb, for the number
of  binary nucleon-nucleon collisions that occur per unit area.
A factor $K\sim 2$ simulates next to leading order corrections.

For symmetric systems, $A+A$, with $G_A\approx A G$.
the inclusive gluon jet production cross section is obtained by integrating
over $y_2$ with $y_1=y$ and $k_\perp$ fixed. 
To about $50\%$ accuracy, the single inclusive gluon rapidity density
in central collisions can be estimated by the following simple pocket 
formula\cite{Gyulassy:1997vt}
\beq
        {{dN} \over {dydt}} \approx  {{A^2 \over {\pi R^2}}
        2 N_g(x_\perp,t)}x_\perp  G(x_\perp,t) 
\left({{d\sigma_{gg}^{el}} \over {dt}}\right)_R\propto A^{4/3}
\eeq{dnint}
where $N_g(x_\perp,t)=\int_{x_\perp}^1 dx G(x,t)$ is the total
 number of ``hard"
gluons coming down the beam pipe in an nuclear area $\pi R^2$ 
that can knock out 
 unsuspecting gluons from the other nucleus.
This copious mini-jet mechanism is believed to the dominant source of the 
 gluon plasma that will be created when
RHIC (finally) begins operation.

\begin{figure}
\vspace{1.in}
\centerline{\epsfxsize= 3.0 in 
\epsfbox{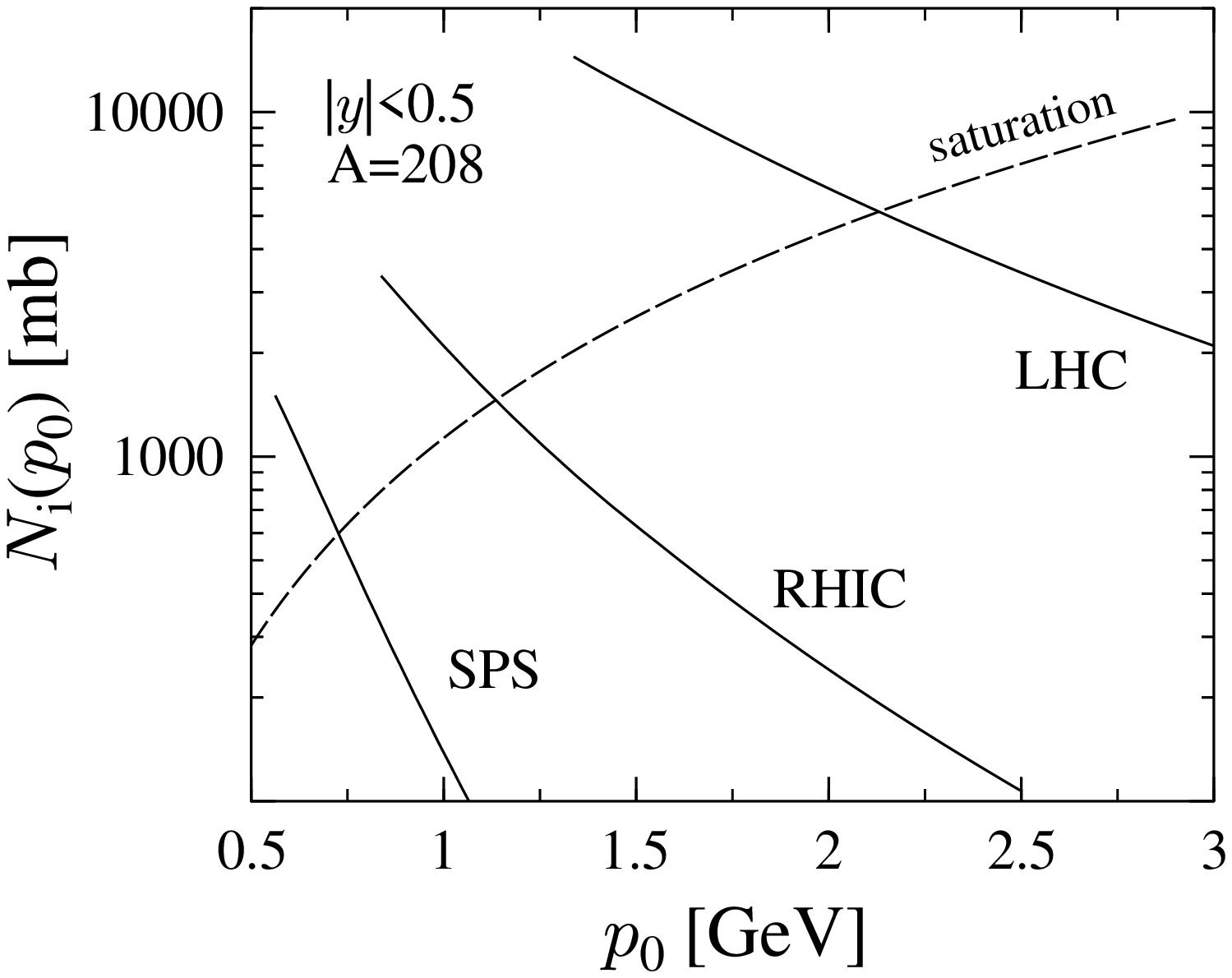}\epsfxsize= 3.0 in 
  \epsfbox{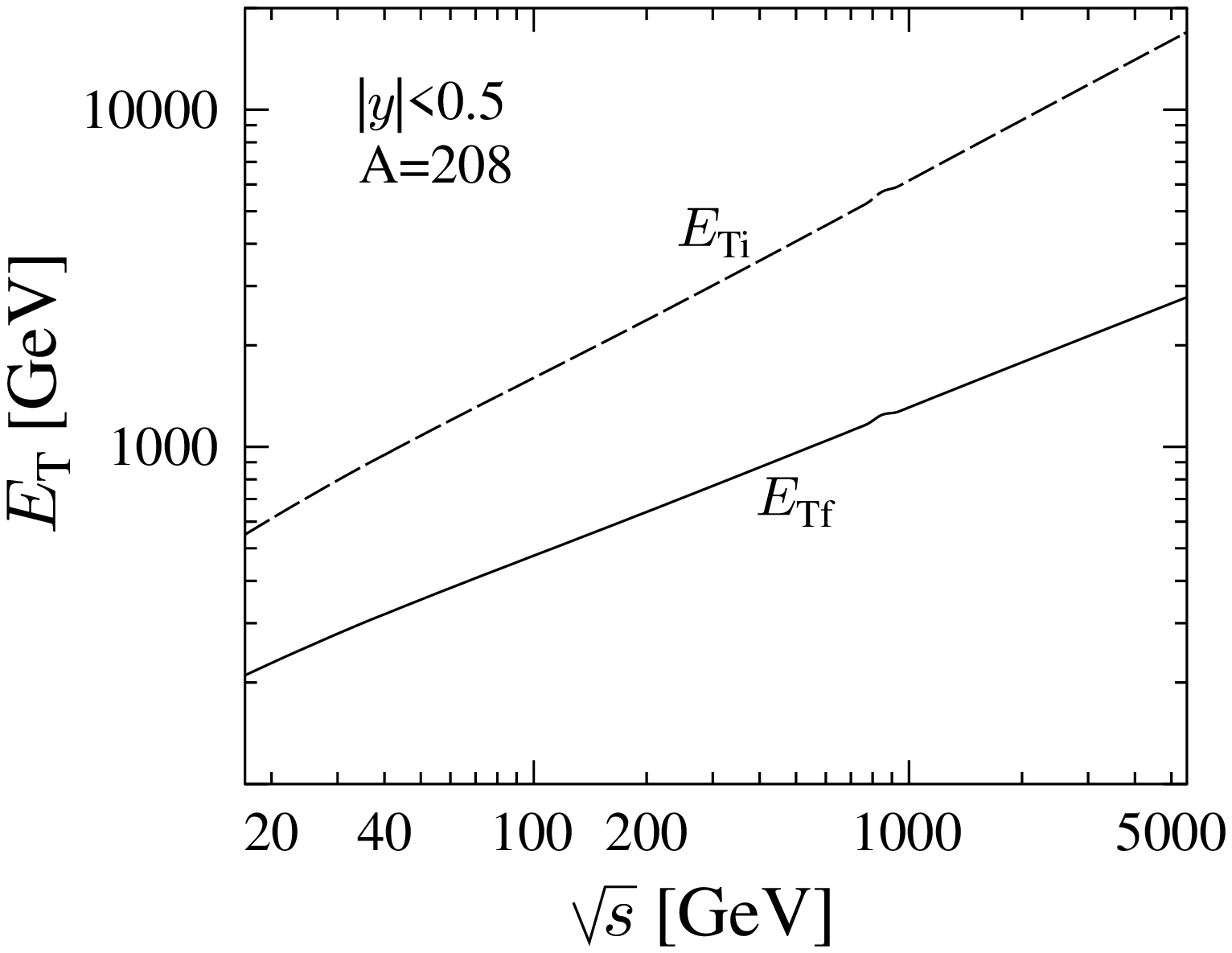}}
\vspace{-2.0in}
\caption{Expected copious mini-jet production in A+A collisions
 from ref. \protect{\cite{Eskola:2000fc}}. The dashed curves
in both figures correspond to an estimated  maximum number (saturation)
at which the nuclear
area if filled with  closed packed mini jet gluons $(N(p_0)=p_0^2R^2)$.
($N_i$ is dimensionless, $[mb]$ is a typo in fig.). The magnitude of possible
hydrodynamic transverse energy loss due to longitudinal work in an
ideal $p=\epsilon/3$ quark gluon plasma is shown by $E_{Tf}$.}
\label{fig:mini}
\end{figure}

Recent upper bound  estimates 
of the total gluon rapidity density as a function of
the CM energy 
from \cite{Eskola:2000fc}
are shown in Fig.(\ref{fig:mini}). 
The differential yields are integrated down to a transverse momentum scale
$p_0\sim 1-2$ GeV. This scale separates the ``soft"
 nonperturbative beam jet fragmentation domain from
the calculable perturbative one above. 
The curve marked saturation\cite{Blaizot:1987nc}
is an upper bound marking the point where the transverse
gluon density of mini-jets becomes so high that the
newly liberated gluons completely fill the nuclear area,
i.e., $ dN/dy \approx p_0^2 R^2$. At that point, higher order
gluon absorption may limit the further increase of
the gluon number. At RHIC energies these
estimates yield up  1500 gluons per unit rapidity.
My more conservative estimates together with
X.N. Wang\cite{Wang:1991ht,Wang:1997yf} gives 500  gluons
per unit rapidity when initial and final state radiation is also taken into
account. This is obtained with a fixed $p_0=2$ GeV,
that was found to be consistent with all available $p\bar{p}$ and low energy
$AA$ data using the HIJING mini-jet event.

\begin{figure}[htb]
\centerline{\hspace{0cm}
{\hbox{\psfig{figure=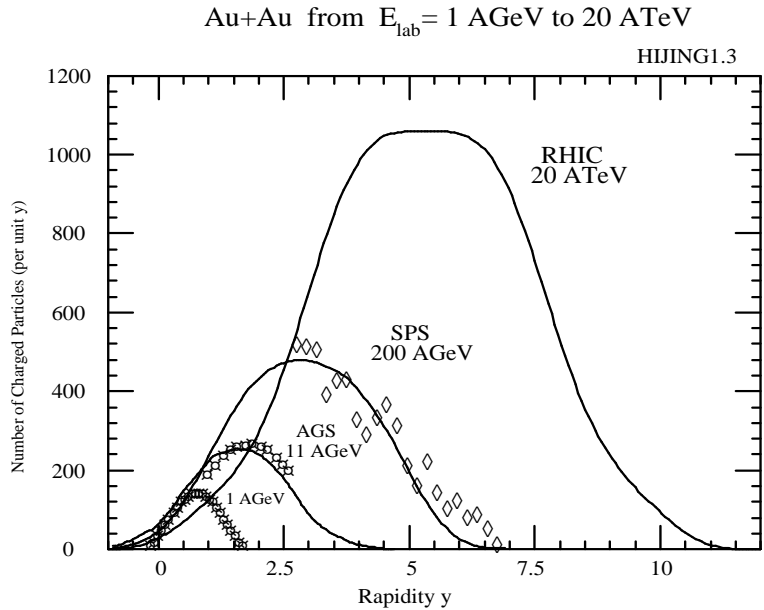,angle=-90,width=10cm}}}
}
\caption{Prediction of the height of "Mt. RHIC"
at $E_{cm}=100$ AGeV based on 
HIJING\protect{\cite{Wang:1991ht}} event
generator from Quark Matter 95\protect{\cite{Gyulassy:1995jz}}. 
Comparison to  measured charged particle rapidity
densities in $Au+Au$ reactions from $E_{lab}=1,10,200$ AGeV is also 
shown.}
\label{fig:mtrhic}
\end{figure}

The energy dependence of the final charged particle 
radidity density from HIJING, including soft beam jet fragmentation,
 is shown in Fig.(\ref{fig:mtrhic}).
Approximately one half of the height ``Mt. RHIC" comes from soft 
beam jet fragmentation processes
modeled in HIJING using Lund/Fritiof strings.
The initial energy density reached in such collisions can be
estimated using the Bjorken formula 
\beq \epsilon(\tau_0) \approx
\frac{1}{\pi R^2\tau_0} \frac{dE_T}{dy} \eeq{bj} For $p_0\sim 1-2$ GeV,
$dE_T/dy\sim 400-2000$ GeV, and so $\epsilon(\tau_0 \sim 0.5\;{\rm fm/c}) > 10$ GeV/fm$^3$ 
should be easily reached,  well inside  the
the deconfinement phase of QCD. 
 At SPS energies, on the other hand, my  estimates indicate that
nuclear collisions may just reach the 
transition region and depart from that region in  a very short time.

Since the the time of this talk, provocative and somewhat  overstated press
releases\cite{cernhype} have been issued from CERN stating that ``the
experiments on CERN's Heavy Ion programme presented compelling evidence for the
existence of a new state of matter in which quarks, instead of being bound up
into more complex particles such as protons and neutrons, are liberated to roam
freely.  ... We now have evidence of a new state of matter where quarks and
gluons are not confined."

As discussed below, I disagree with the above interpretation of that truly
impressive body of data. What is compelling is that {\em some} form of matter,
much denser than ever studied before, was created.  Inferences about quark and
gluon degrees of freedom are  based on qualitative scenarios
and schematic models.   At the relatively low momentum scales accessible
at SPS energies, the quarks and gluon degrees of freedom in the dense matter
are mostly not resolvable. Even at the highest transverse momentum, the pion
spectra were shown to be very sensitive to nonperturbative model assumptions
such as to the magnitude of intrinsic momenta
and soft initial state interactions\cite{Gyulassy:1998nc}. The
dynamics of the  non-perturbative beam jet fragmentation and hadronic final
state interactions cannot be disentangled at SPS\cite{Bass:1999vz}.  It is
useful to recall that in experiments on $p\bar{p}$ scattering, it was only
possibly to see unambiguous evidence for the tell-tale Rutherford scattering of
point-like partons when collider energies $\sqrt{s}>200-2000$ GeV became
available. At RHIC high $p_\perp > 10$ GeV probes become kinematically
available and hence very small wavelength resolution of partonic degrees of
freedom finally becomes kinematically possible.  While there is an abundance of
interesting signatures showing that dense matter was formed at the SPS
(through the non-linear in dependence of several observables on multiplicity or
$A$), the bottom line is that those data have said nothing about whether the
QCD predictions in Figs 1-3 are correct or not.  We simply need higher
resolution. RHIC, with its factor of ten increase in C.M. energy reaches
a factor of  ten deeper into the new phase reaches. Coupled with
the availability of ten times shorter wavelength probes, it
should finally become possible to actually see
direct evidence of ``freely roaming quarks and gluons".

\subsection{Global Signatures of  Collective Dynamics}

The simplest but often ignored global barometer
of collectivity in nuclear reactions is the A and energy dependence
of the transverse energy and charged particle
 rapidity density. At RHIC energies, HIJING
predicts that initial transverse energy density 
in central $A+A$ collisions
scales nonlinearly with A 
\beq
\frac{dE_\perp}{dy} \approx 1 \;{\rm GeV}\; A^{1.3}\;(1+ \log\frac{\sqrt{s}}{200})
\eeq{et}
This leads to about $1$ TeV per unit rapidity. In
\cite{Eskola:2000fc}, on the other hand, it was found that gluon saturation
could limit the A dependence to approximately linear, $E_\perp\sim
A^{1.04}$, but with a value several times that of HIJING.
In Fig.(\ref{fig:mini}), the initial gluon density actually grows
less than linear $A^{0.92}$ in that model.
The $A^{1.3}$ scaling of HIJING with
its conservative  $p_0=2$ GeV scale fixed by $p\bar{p}$ data
is simply due to the number of binary interactions via 
(\ref{dnint}).
The initial energy density in HIJING varies approximately as
\beq
\epsilon_0\approx 0.6\; {\rm GeV/fm}^3\; A^{0.63}\;(1+ \log\frac{\sqrt{s}}{200})
\; \; .\eeq{ep0}
In \cite{Eskola:2000fc} its magnitude and scaling
are predicted to go as $\epsilon(\tau=1/p_{sat})\approx 0.1 A^{0.5}
s^{0.38}$. 

If local equilibrium is achieved and maintained, then a very basic prediction 
of hydrodynamics
is that   longitudinal boost invariant  expansion
together with $pdV$ work done by pushing matter down the beam pipe
will cool the plasma and convert some its random transverse
energy into collective longitudinal kinetic energy.
 For an equation of state, $p=c_s^2 \epsilon$,
 this cooling and expansion causes the energy
 density to decrease with proper time as
\beq
\epsilon(\tau)=\epsilon(\tau_0)\left(\frac{\tau_0}{\tau}\right)^{1+c_s^2}
\eeq{bj}
Entropy conservation leads to a conservation
of $\tau n(\tau)$, where $n$ is the proper parton density. 
At least if the matter is initially deep in the plasma phase, then
(as seen in Fig.3) longitudinal  will be done with $c_s^2\approx 1/3$.
Consequently, the transverse energy per particle
should decrease by a factor 2-3 before freeze-out as\cite{Nayak:2000js,Eskola:2000fc}
\beq
e_\perp(\tau)=
\frac{dE_\perp}{dN}=e_\perp(\tau_0)\left(\frac{\tau_0}{\tau}\right)^{c_s^2}
\eeq{work} 
However,  dissipative effects due to finite
mean free paths reduce the effective pressure in any system.
For the Bjorken expansion, the relaxation time,
$\tau_c=1/(\sigma_{T}n)\propto \tau$, 
then increases with time as $n$ decreases.
Numerical solution
of 3+1 D transport equations with pQCD cross sections,
$\sigma_T\sim 2$ mb, indicate that
dissipation reduces the transverse energy loss
for HIJING initial conditions rather significantly
(see detailed comparisons in \cite{Gyulassy:1997zn}).
Ideal hydrodynamics predicts that about one 
half of the initial produced transverse energy
goes into longitudinal work

It is important to keep in mind 
that the commonly assumed freeze-out
prescription with $\tau_f$ fixed on a fixed freeze-out
energy density hypersurface,
$\epsilon(\tau_f)=\epsilon_f\approx 0.15$ GeV/fm$^3$ is only a rough
prescription that is never  accurate\cite{molnar}.
As the interaction size decreases ($A$ or multiplicity decreases),
the effective speed of sound due to dissipation
decreases and the system freezes out earlier. 
A detailed  study of the A or multiplicity
dependence of $dE_\perp/dy$ and $e_T$ is  needed to calibrate
the interplay of pdV work and  dissipation on these barometers.

\begin{figure}
\vspace{0.in}
\centerline{\epsfxsize= 3.0 in 
\epsfbox{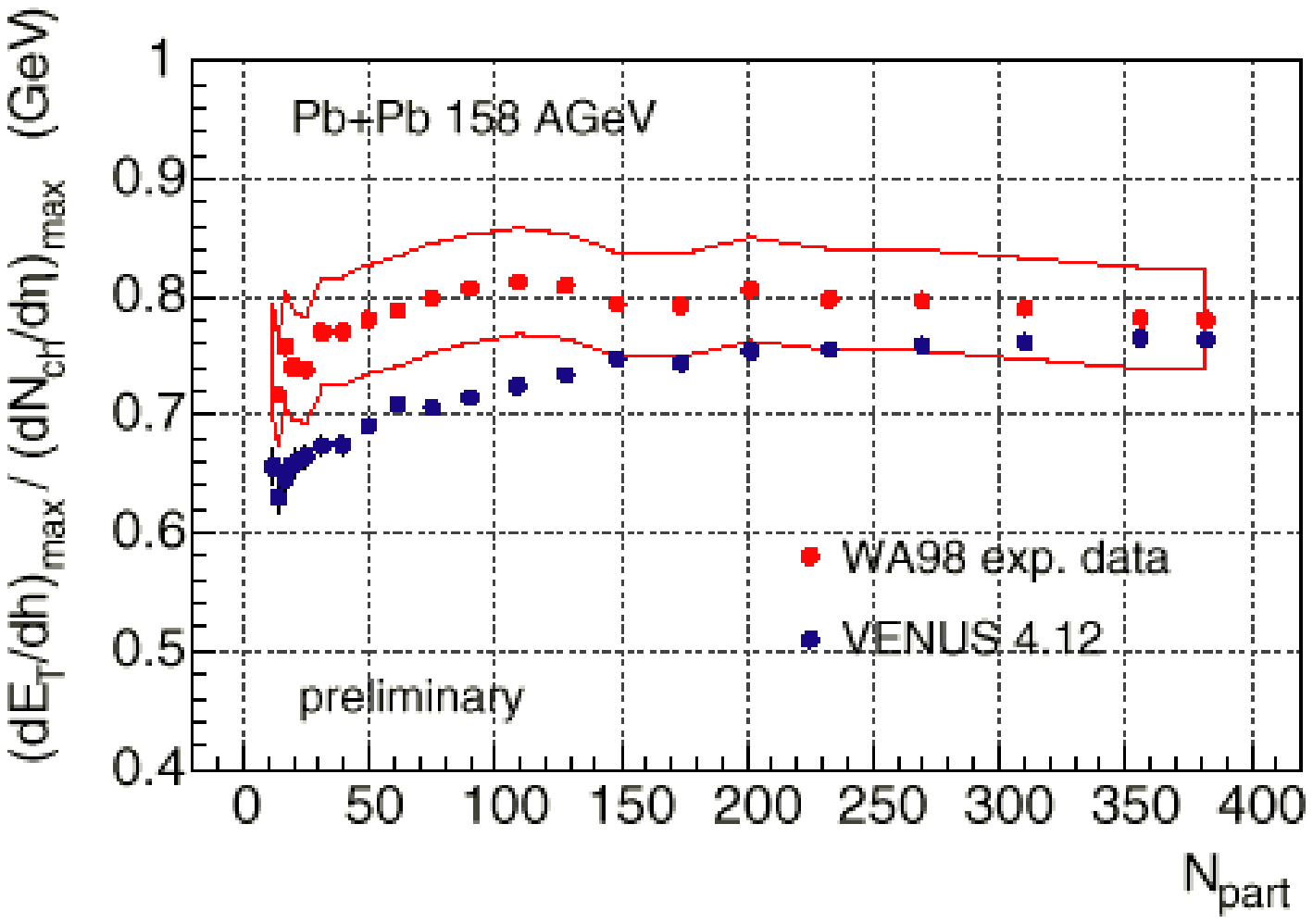}\epsfxsize= 3.0 in 
  \epsfbox{ 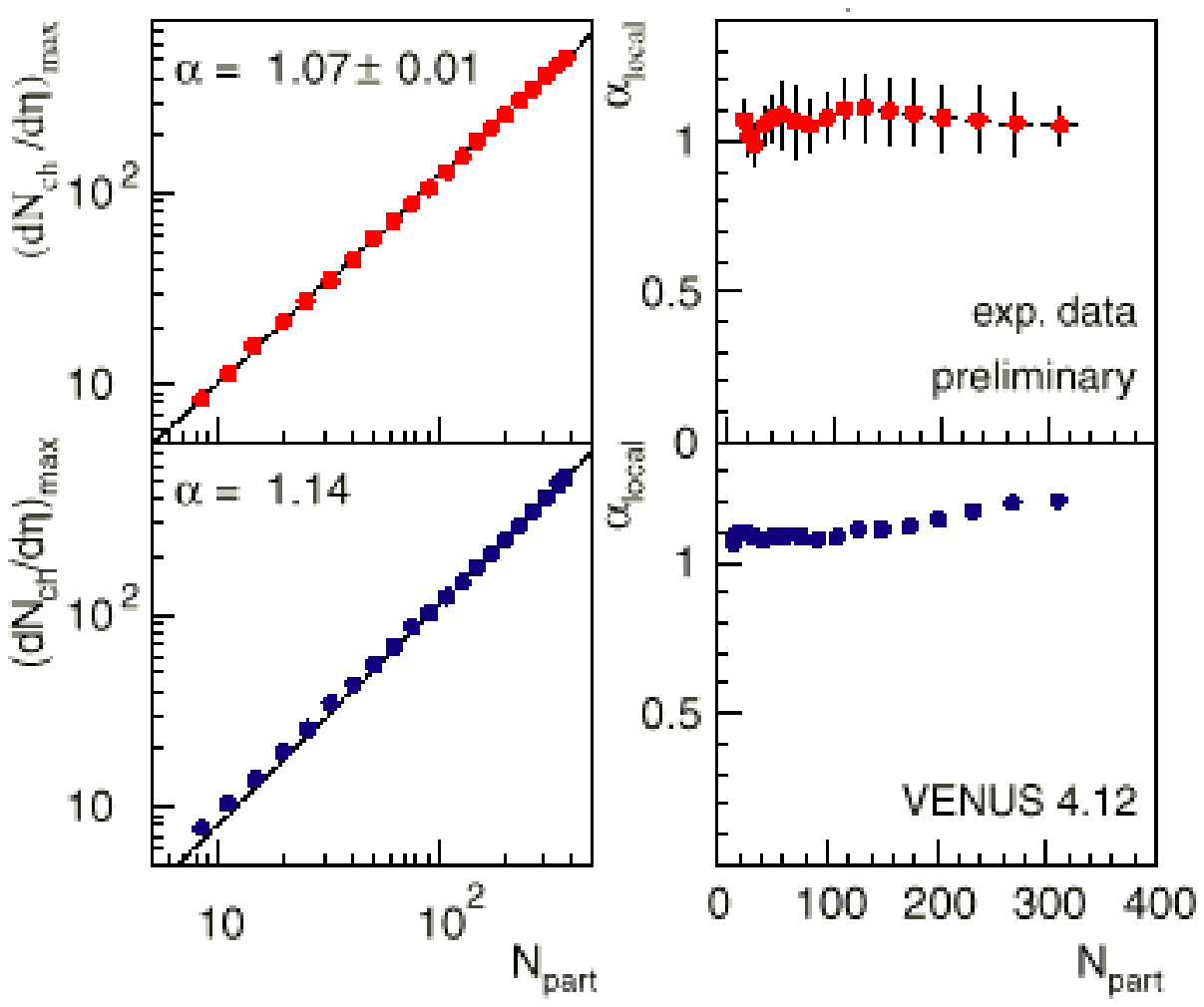}}
\caption{Scaling of transverse energy per charged particle
and and the charged particle radidity density with number of participants
in $Pb+Pb$ from H. Schlagheck et al WA98\protect{\cite{Schlagheck:2000aq}}.
The nearly linear scaling in the number of participants can be interpreted as
no longitudinal work was done by the lazy dense matter at the SPS.}
\label{fig:wa98}
\end{figure}

One of the  major experimental 
discoveries  of WA98, NA49 and the other
SPS experiments  is that
at those energies $dE_\perp/dy$ as well as $dN/dy$ scale with $A$
or wounded nucleon number
 nearly linearly \cite{Schlagheck:2000aq} as shown in Fig.(\ref{fig:wa98}).
Both the $E_\perp$ and the charge multiplicity  increase as
$\sim A^{1.07}$.  These findings differ from the  VENUS
model, which has considerable nonlinearly due to the assumed sea string
contributions.  
In fact, simple Glauber wounded nucleon models reproduce very
well the nearly linear correlation between $E_\perp$ 
and the veto calorimeter (spectator) energy 
observed in all experiments at SPS\cite{Kharzeev:1997yx}. 
The implications of these data depend on the $A$ dependence
of the initial conditions of course. One view\cite{Eskola:2000fc} is that
the initial $e_T$ can scale arbitrarily  with $A$, 
but because perfect local equilibrium is maintained up to a critical sharp
freeze-out hypersurface, the final $E_T$ and $e_T$ always
scale linearly with $A$.
My view 
is that at the SPS,  pQCD is mostly inapplicable to bulk phenomena 
 and the linear dependence
arises from  additive nature of soft beam fragmentation together with 
the {\em absence} of $pdV$ work at early times.
If the QGP transition region is just barely
reached, as I believe without experimental proof,
then the softness of the QCD equation of state with
 $c_s^2\ll 1/3$ seen in Fig.(\ref{fig:eos}) and dissipation
can conspire to prevent the dense matter from
performing longitudinal work.
However, it is impossible to tell from the data 
whether the observed null effect  is then  due
to a low pressure  Hagedorn resonance gas of hadrons or 
to a low pressure lazy 
``plasma'' with $c_s\ll 1$.
One has to go to higher energies to the plasma a chance to work.

\begin{figure}
\vspace{0.in}
\centerline{\epsfxsize= 3.0 in 
\epsfbox{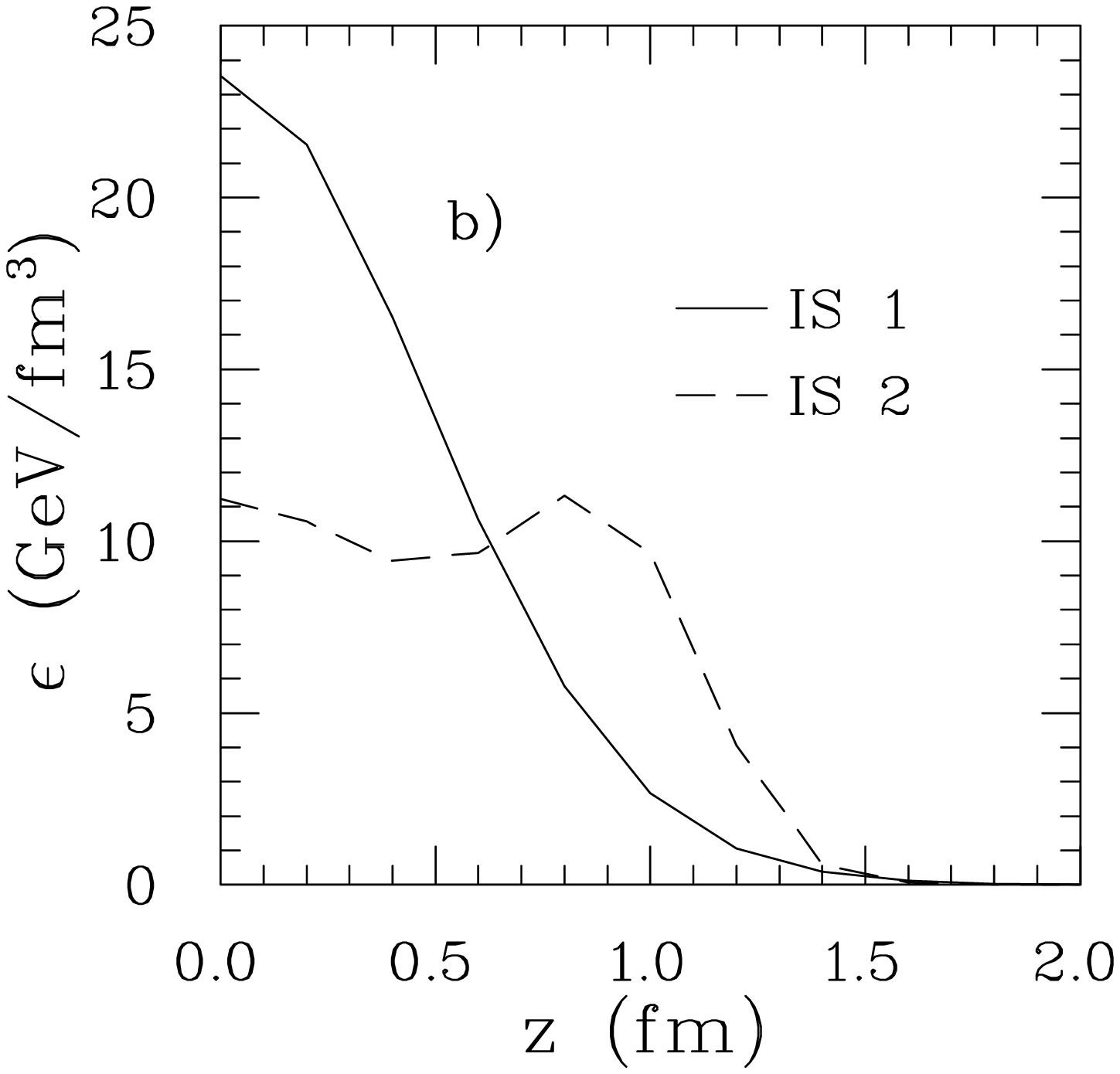}\epsfxsize= 3.0 in 
  \epsfbox{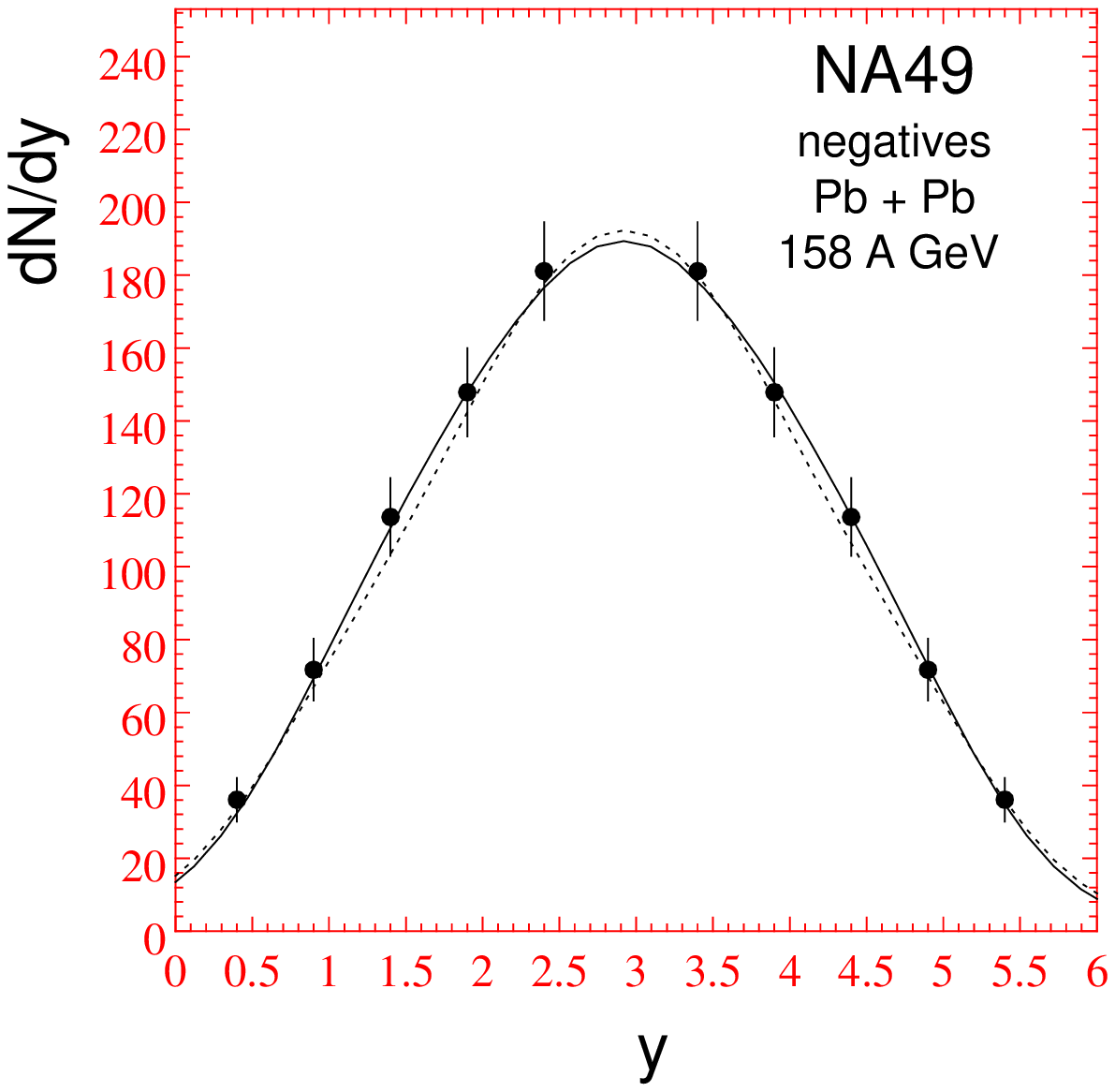}}
\caption{Some different initial conditions that lead with suitable
initial velocity fields via ideal hydrodynamics
to same
final observable  pion rapidity
density at SPS from ref. \protect{\cite{Huovinen:1999tq}}.}
\label{fig:ruus}
\end{figure}

In spite of the above remarks, ideal Euler  hydrodynamic
calculations  have  in fact been
 able to fit the null effect in the data quite well. This is possible
by assuming arbitrary 
 initial conditions chosen such that
that after hydro does its work, the final results just happens
to reproduce the data. 
An illustration of this degree of
 freedom  is shown in Fig.(\ref{fig:ruus}) from  \cite{Huovinen:1999tq}.
By adjusting the initial four velocity field appropriately,
the calculated final pion distributions can be made to reproduce
 the data starting from
two completely different initial energy density profiles.
The same good fit to data can be obtained
also for {\em ANY} assumed equation of state of dense matter
simply by readjusting the initial conditions
appropriately. Therefore, it is not surprising
 that  even hand calculator fireball models are able to fit much of the
data {\em without} invoking any dynamical or thermodynamical assumptions
about the properties of  dense matter prior to freeze-out
\cite{Heinz:1999kb,Ster:1999ib}.
It is clear that  inferring
the existence of freely roaming quarks and 
gluons based on such fits is impossible
\cite{Heinz:2000bk,cernhype}. 

In order to use the global barometers
to search for evidence of collective phenomena in dense matter at RHIC,
it is first necessary to eliminate  the freedom to choose arbitrary
initial conditions\cite{Dumitru:1999es}. At collider energies,
 pQCD \cite{Eskola:2000fc,Eskola:1989yh,Blaizot:1987nc,Gyulassy:1997vt}
and  non-Abelian field techniques
\cite{McLerran:1994ni,Krasnitz:1999wc}
 provide the needed theoretical calibration tools to fix initial conditions
 At RHIC and LHC, the initial plasma is expected in any case to be
so deep in the deconfined phase that the soft $p\ll\epsilon/3$ transition
region and dissipation cannot spoil the longitudinal
work as is the case at SPS. 
From detailed covariant transport calculations\cite{Gyulassy:1997zn,molnar}
the barometric evidence for longitudinal
work should finally be observable in spite of
finite mean free path effects. Deconvolution of the equation of state and
dissipative corrections requires however
a detailed study  of the $A$ and multiplicity dependence of the global $E_T$
barometer.

This is the first important 
signatures to look forward to at RHIC.
While it is not possible to converge  on 
 why  no longitudinal
work was observed at the SPS, 
a positive signature at RHIC would render such discussion  mute.
We can only be sure that a plasma was created if it does something
collective!
The global $E_T$ and multiplicity systematics as a function
of centrality as well as $A$ therefore provide
 key handles in this  search.

\section{Transverse Flow}

\begin{figure}
\vspace{0.in}
\centerline{\epsfxsize= 2.5 in 
\epsfbox{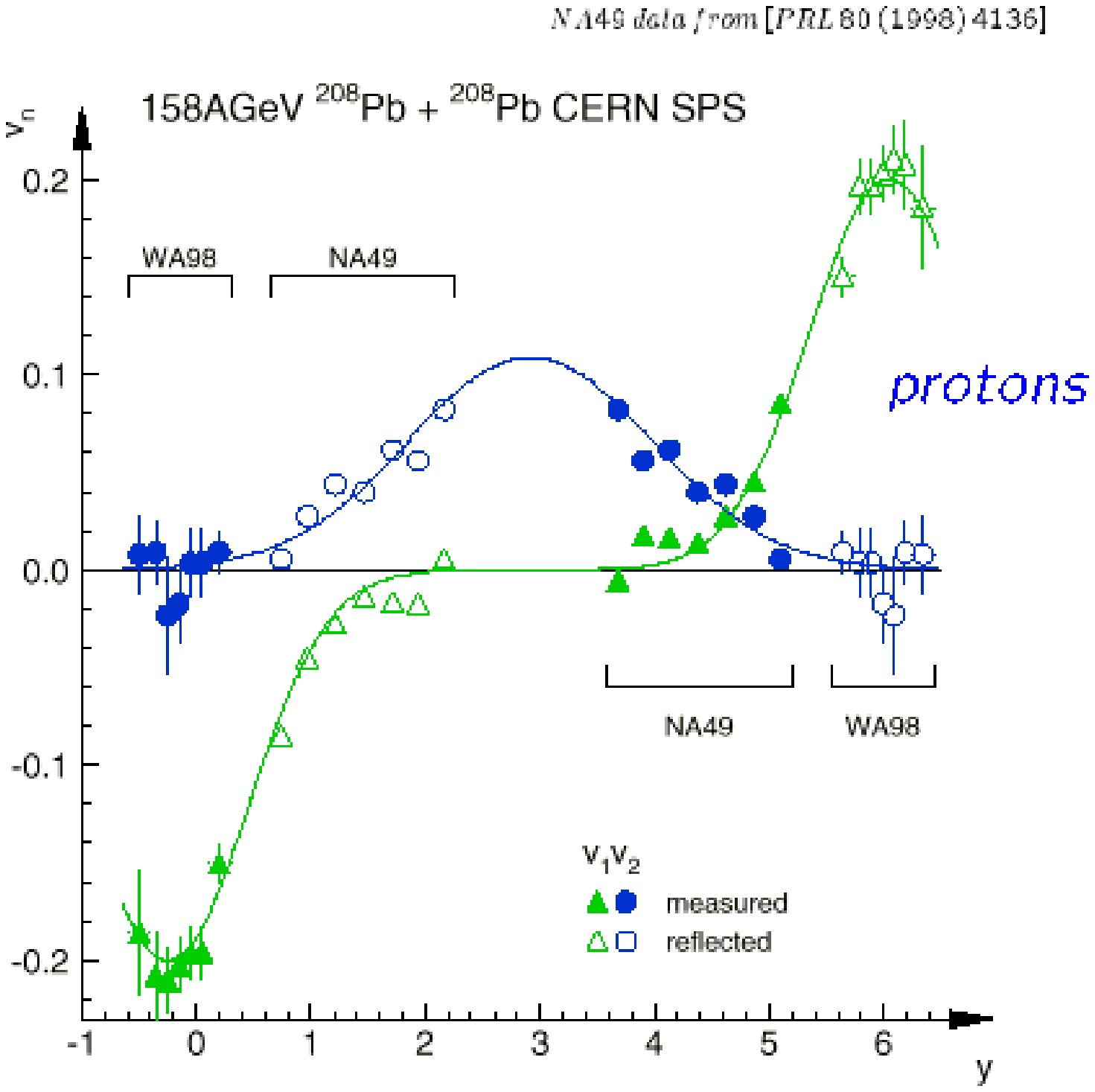}
\epsfxsize= 3.0 in 
\hspace{0.3in}{\hbox{\psfig{figure=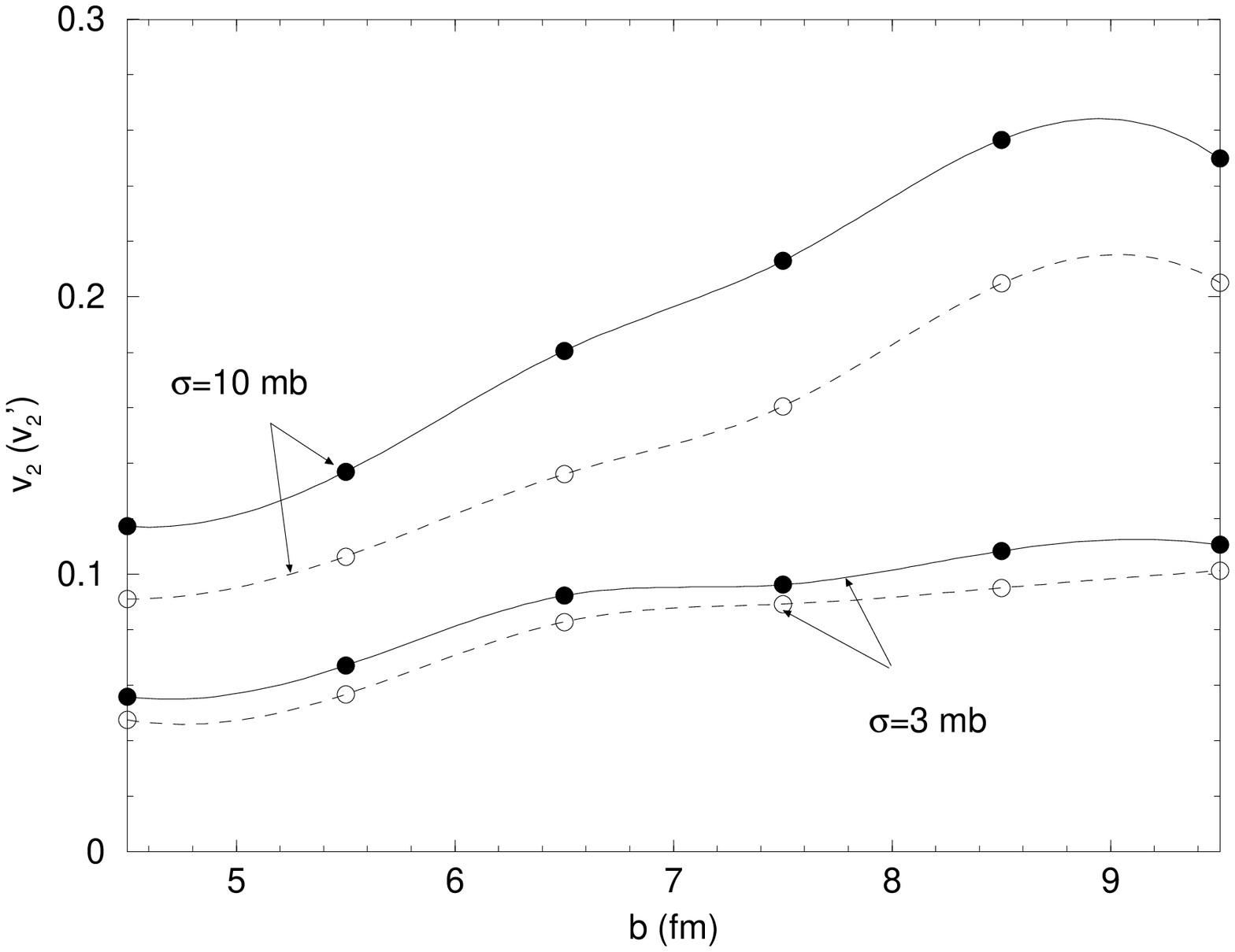,angle=-90,width=6cm}}}}
\caption{(left) Azimuthal flow analysis
in $Pb+Pb$ from H. Schlagheck et al WA98
\protect{\cite{Schlagheck:2000aq,wa98flow}} 
and NA49 data\protect{\cite{Voloshin:2000gs}}. 
(Right) Predicted dependence of $v_2$ at mid rapidity $Au+Au$ at
RHIC from parton cascade calculations 
ZPC\protect{\cite{Zhang:1999rs}} with HIJING initial conditions.}
\label{fig:wa98flow}
\end{figure}

In contrast to global transverse energy barometer a completely
different measure of barometric collectivity is afforded by the study
of the triple differential distributions, $dN/dyd^2\vp_\perp$. Already
at sub-luminal Bevalac energies ($<1$ AGeV), azimuthally asymmetric
 collective directed and elliptic flow were
discovered long ago. For non central collisions, $b\ne0$, the asymmetric
transverse coordinate profile of the reaction region 
 leads to different gradients of the pressure as a function of
the azimuthal angle relative to beam axis.
  This leads to a ``bounce'' off of projectile and
target fragments in the reaction plane and to azimuthally asymmetric
transverse momentum dependence of particles with short mean free at
mid rapidities.  This phenomenon has now been observed at both
AGS and SPS energies as well.  It will certainly be there also at RHIC and
LHC. In Fig.(\ref{fig:wa98flow}) the first two Fourier components of the
azimuthal flow patterns are shown:
\beq
\frac{dN}{dyd^2\vp_\perp}= v_0 (1+ 2v_1 \cos( \phi-\phi_R)+
2v_2 \cos( 2(\phi-\phi_R)) +\cdots)
\eeq{v2}
Azimuthal asymmetric collectivity
is clearly observed at the SPS. The important  question is how this
type of barometer could serve to help search for evidence
of the QCD transition. Unlike the global $E_\perp$ barometer
discussed in the previous section,
transverse flow can develop at later times 
because the gradients are controlled by
the transverse size of the nucleus,
not the proper time interval relative to formation.
In \cite{jolli1} the idea was proposed that one could use $v_2$
for example to study the predicted softening of the QCD equation
of state. 
Typically, hydrodynamics calculations lead to a factor of two smaller $v_2$ for
an equation of state with a soft critical point as in Fig 3 versus
one in which the speed of sound remains 1$/\sqrt{3}$.
Searches for anomalous $v_2$ dependence as well as $v_1$ are
underway\cite{wreis1}. As with the global barometer, dissipation
can of course also simulate a soft equation of state. 
In ref\cite{Zhang:1999rs} we studied the dependence
of $v_2$ on the transport parton cross section. The results
 shown in Fig(\ref{fig:wa98flow})
 for HIJING initial conditions indicate that
 there is indeed a significant reduction of $v_2$ relative to hydrodynamics,
 but that the impact
parameter (multiplicity) dependence of that observable
can again be used to disentangle the equation of state versus viscosity
effects as in the case of the global $E_T$ barometer.

\section{The QGP Stall and Time delay}

\begin{figure}
\vspace{0.in}
%
\hspace{1.in}
{\hbox{\psfig{figure=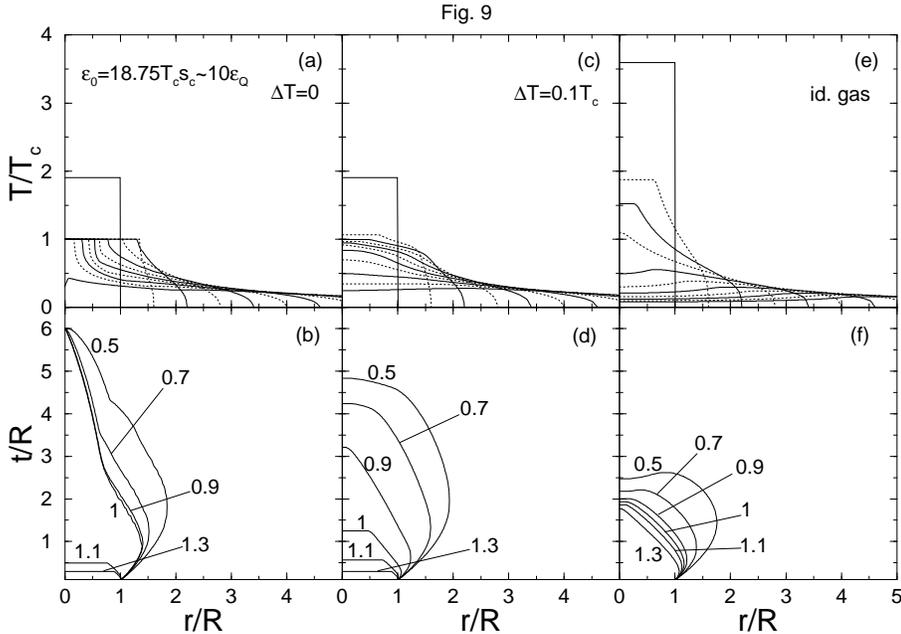,angle=-90,width=7cm}}}
\vspace{-1in}
\caption{Evolution of isothermal contours for
a boost invariant QGP cylinder 
with initial energy density $\sim 20$ GeV/fm$^3$
for different equations of state from ref.
\protect{\cite{Rischke:1996em}}.
For a strong first order transition the
QGP "stall" is very strong, but it also remains
significant even if the transition is only a rapid cross over 
consistent with Fig.3. For an ideal gas with  $p=\epsilon/3$
on the other hand, the freeze-out time scale and radius are  approximately
equal. An experimental probe of this decoupling geometry is provided
by pion or kaon interferometry.}
\label{fig:hbt}
\end{figure}

Hadron interferometry has been developed into a fine art
in heavy ion collisions to image the space-time region of
the decoupling 4-volume. In \cite{pratt} it was proposed that
a possible  signature of the QGP phase transition
would be a time delay associated with very slow
hadronization. The plasma "burns" into hadronic ashes along
deflagration front that moves very slowly if the entropy drop across the
transition is large\cite{vanH}.
Fig.\ref{fig:hbt} shows the evolution of a Bjorken
cylinder with time and transverse coordinate from a
hydrodynamic simulation with different equations of state
from \cite{Rischke:1996em}.
The main point to note is that time delay is a robust generic
signature of a rapid cross over transition  of the entropy density.
In particular the "stall" is expected  even for a smooth
cross-over transition 
as long as the width $\Delta T/T_c <0.1$. However, its magnitude also
depends on the entropy drop across that region. The figures are for
an entropy drop by a factor 10 consistent with the lattice QCD results.
 Unfortunately, as noted before
lattice QCD has not yet resolved the hadronic world below $T_c$
due to numerical problems. If the entropy jump is much smaller,
then this signature would also disappear.

High statistics measurements of pion and 
kaon interferometry searches for time
delay at AGS and SPS have come up empty handed thus far.  No time delay 
has ever
been observed in any nuclear reactions thus far. 
This could be due to (a) the absence of
a large rapid entropy drop in real QCD, or (b)
to unfavorable kinematic conditions at AGS and SPS energies.  From the
hydrodynamic calculations in \cite{Rischke:1996em},
 it was found that a large 
time delay signal requires that the initial energy density be deep
within the plasma phase.
The slow deflagration waves seen in  Fig.\ref{fig:hbt}
do not arise  if the initial energy density is
close to the transition region because the initial longitudinal
expansion cools the plasma too rapidly. 
The optimal conditions to see this effect
was  predicted\cite{Rischke:1996em} to occur  at RHIC energies.
 For much higher energies
(LHC), on the other hand, the transverse expansion of the plasma 
has too much time to develop and that spoils 
the possibility of a slowly burning plasma stall.
I note that further work
\cite{Bernard:1997bq} has shown that high $p_\perp$ kaon interferometry 
adds an especially sensitivity handle in the search for this 
time delay signature. If ever observed, the time delay signature
would be smoking gun that a new form of matter with bulk collective properties 
was created.

\section{The J/Psi puzzle}

In 1986, Matsui and Satz proposed an intriguing direct measure 
of the transmutation of the $q\bar{q}$ forces in Fig. 1.
The idea was that $J/\psi$ can form in the vacuum because
the confining Luscher potential can bind a $c\bar{c}$ pair into
that vector meson. If that pair were placed in a hot medium in which
the 
chromo-Debye screening potential is short ranged, 
then above the temperature where the screening length is
smaller than the $J/\psi$ radius, the $c\bar{c}$ would 
become unbound and and the charm quarks would emerge from the reaction region
as an open charm $D\bar{D}$ pair. They thus predicted that $J/\psi$ suppression
would be a smoking gun for the deconfinement transition.

\begin{figure}
%
\centerline{\epsfxsize= 2.5 in \epsfbox{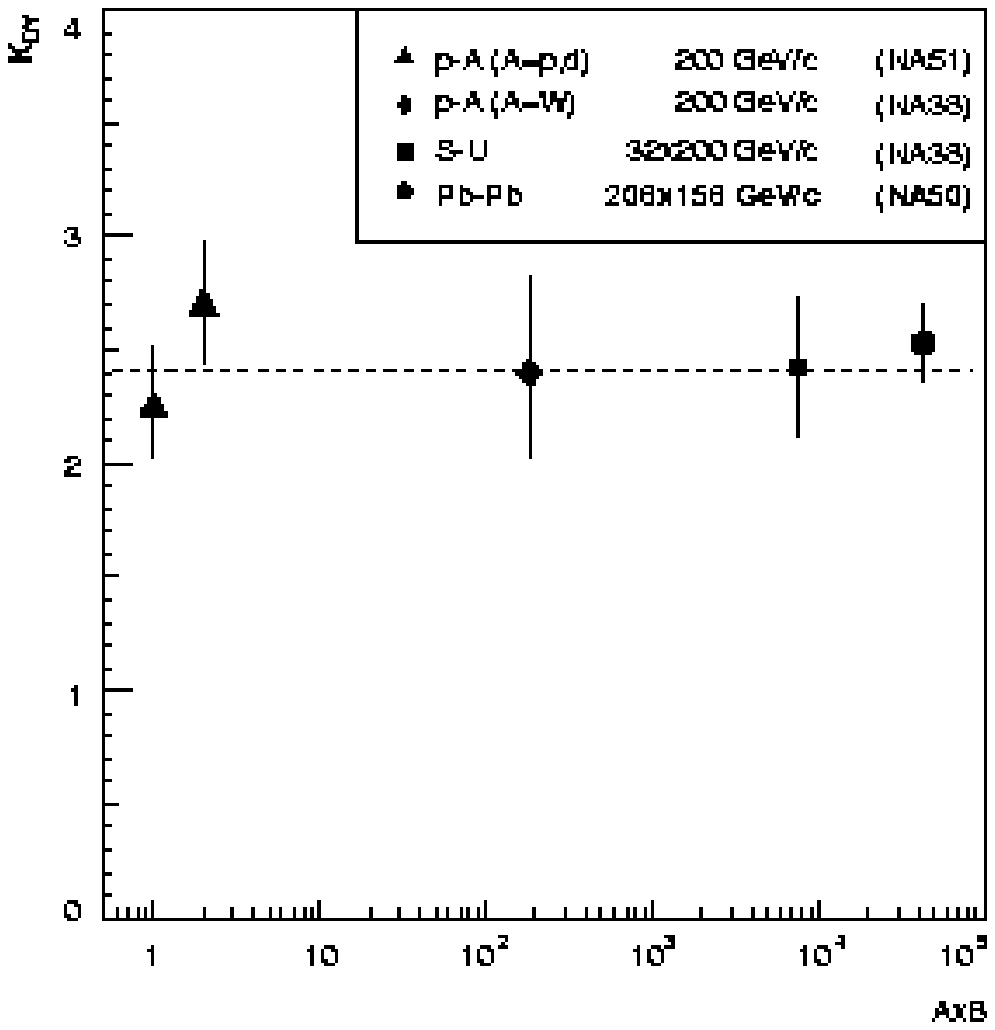}
\epsfxsize= 2.5 in 
  \epsfbox{ 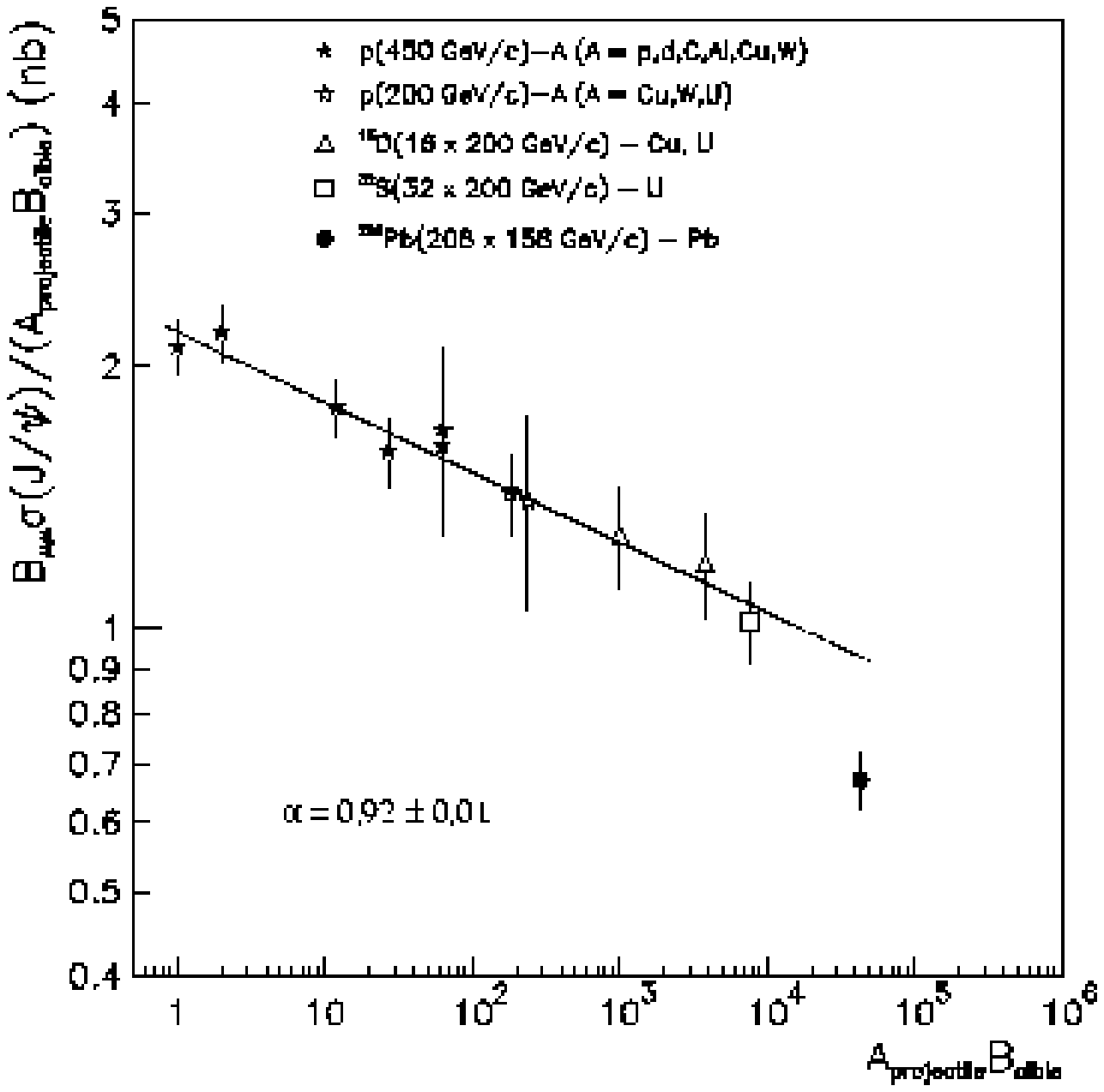}
}
\caption{Drell-Yan scales as $(AB)^1$ very accurately from NA38, NA50.
However, $J/\psi$ scale as $(AB)^{0.9}$ showing
that their production is  suppressed
with increasing nuclear volume  even in $p+A$ reactions
Ref.\protect{\cite{Abreu:2000ni}}. The "anomalous" suppression in $Pb+Pb$
is the 25\% deviation from the empirical $(AB)^{0.9}$ expectation.
}
\label{fig:dyvsj}
\end{figure}

$J/\psi$ suppression was first seen  in 1987 in $O+U$
 reactions by NA38. Since then this smoking gun has
(unfortunately) never stopped smoking! $J/\psi$ suppression seems to
be as ubiquitous
as the Yukawa potential. It is now clearly observed in $p+A$ reactions
as seen in Fig.(\ref{fig:dyvsj}). High mass Drell-Yan pairs, 
on the other hand, formed via
 $q\bar{q}\rightarrow \ell \bar{\ell}$
was observed to
scale  perfectly linearly with the number of binary collisions.
This is because 
lepton pairs suffer no final state interactions
and the quark initial state (Cronin) interactions are invisible
in $p_\perp$ integrated DY yields.

\begin{figure}
%
\centerline{\epsfxsize= 2.5 in \epsfbox{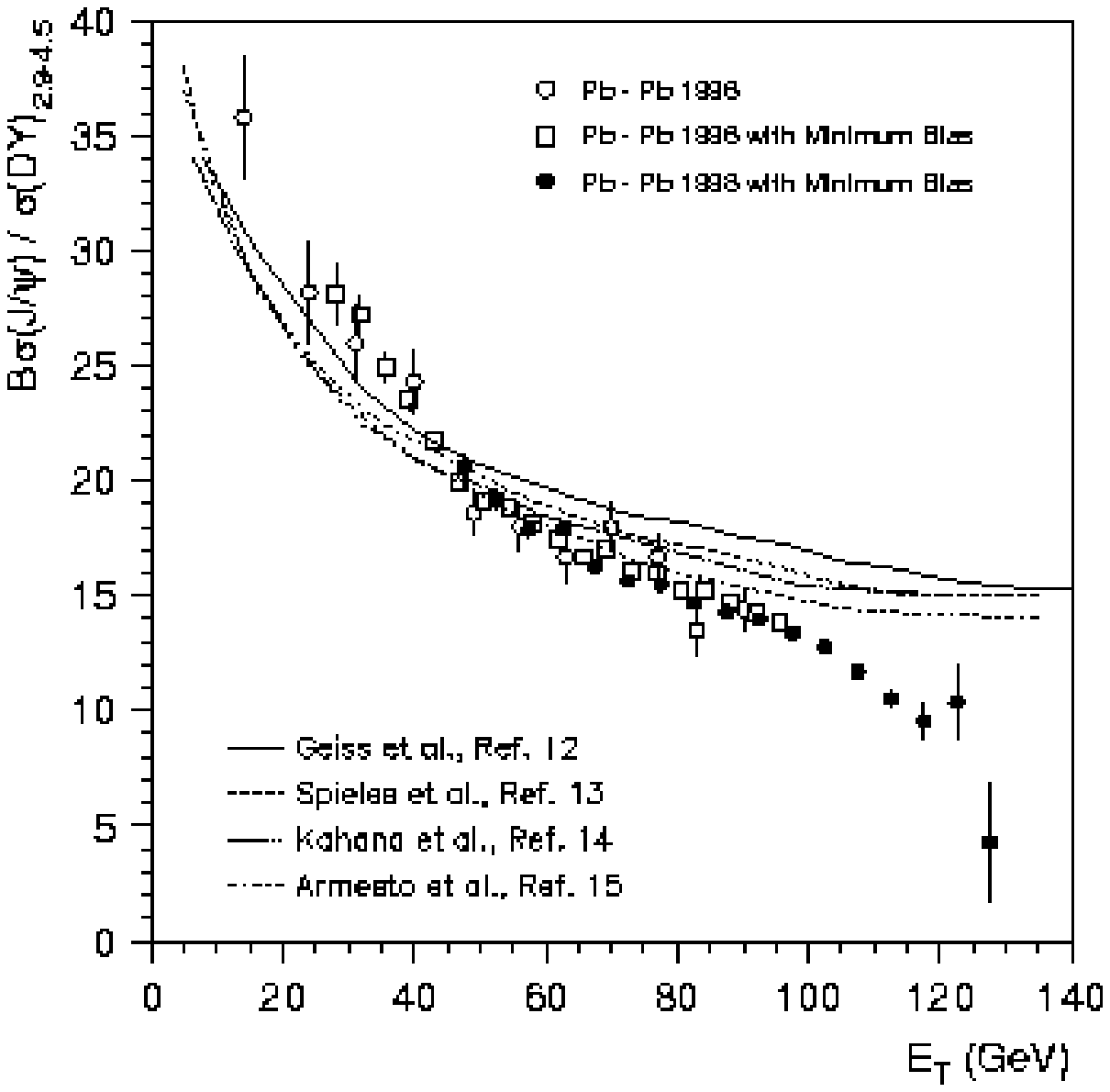}
\epsfxsize= 2.5 in 
  \epsfbox{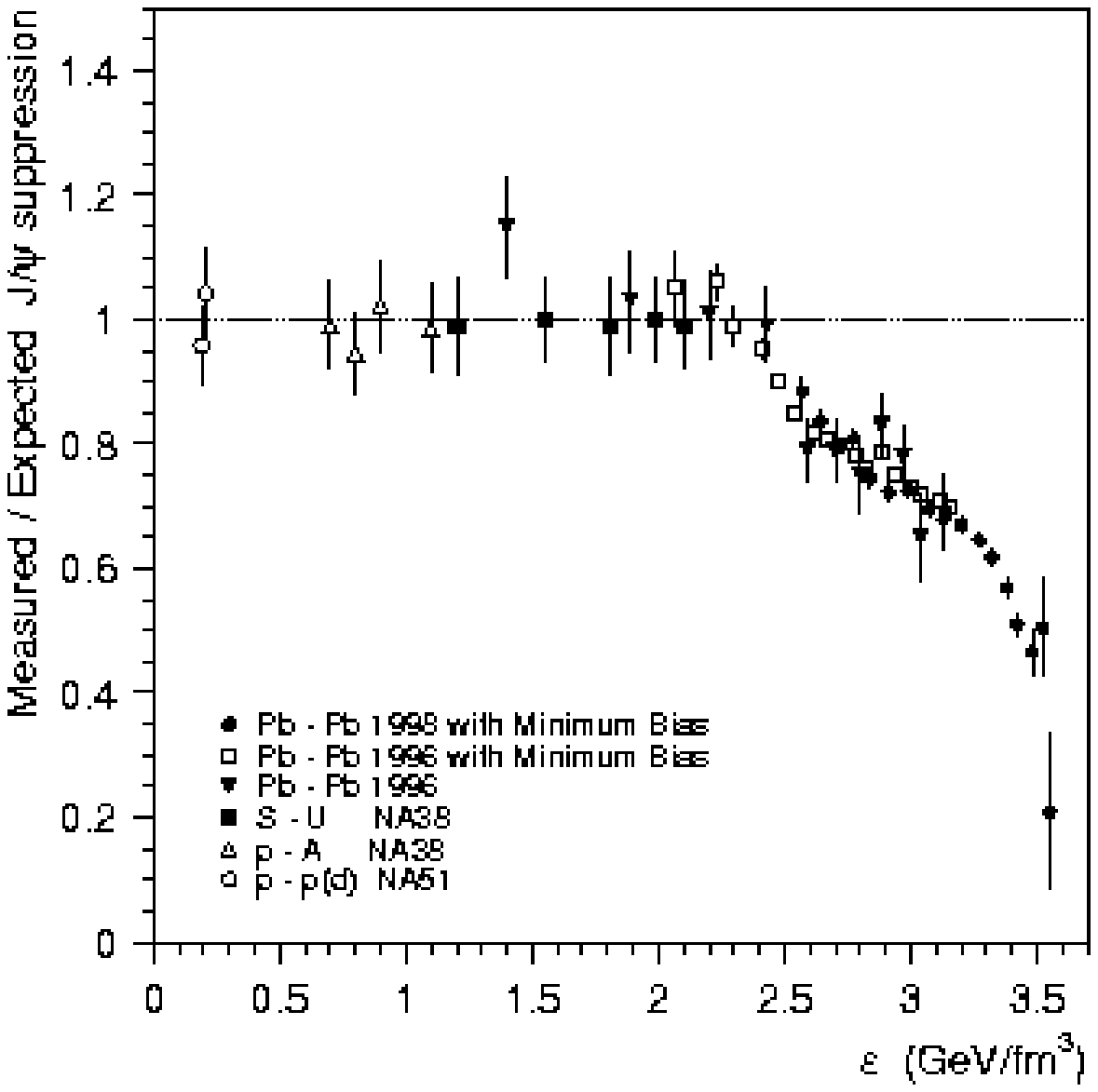}
}
\caption{``Evidence for deconfinement of quarks and gluons from the J/psi
suppression pattern measured in Pb Pb collisions at the CERN-SPS"
claimed by NA50 in Ref.\protect{\cite{Abreu:2000ni}}.
The curves on left show transport theory estimates for hadronic final state 
dissociation. The enhanced suppression relative to $\exp(-\sigma_N \rho_0 L)$
nuclear suppression observed in $p+A$ and $S+U$ is shown versus
a rough estimate of the initial energy density (this scale is uncertain
to at least a factor of two for each point!).}
\label{fig:psina50}
\end{figure}

$J/\psi$ are  suppressed on the other hand as $(AB)^{0.9}$
due to some nuclear effect that is independent of QGP production. In the recent
Pb+Pb analysis an excess 25\% suppression of $J/\psi$ was observed.
 NA50  claims that this enhanced suppression relative
to the $(AB)^{0.9}$ trend from lighter projectile $p,O,S+U$ data
is finally the real smoking gun\cite{Abreu:2000ni}.
From the suppression pattern
in Pb as observed  as a function of centrality (see Fig.(\ref{fig:psina50})),
NA50 has now claimed that in fact ``Together with the results
previously established by the NA38 and NA50 collaborations, a rather clear
picture emerges, indicating a step-wise pattern, with no visible saturation in
the collisions generating the highest energy densities and temperatures. Our
observations exclude the presently available models of $J/\psi$ suppression
based on the absorption of the $J/\psi$ mesons by interactions with the
surrounding hadronic (confined) matter.  The first anomalous step can be
understood as due to the disappearance of the $\chi$ mesons, responsible for a
fraction of the observed $J/\psi$ yield through its (experimentally
unidentified) radiative decay. In proton induced collisions this fraction is
around 30-40\%.  The second drop signals the presence of energy densities high
enough to also dissolve the more tightly bound $J/\psi$ charmonium state."

\begin{figure}
\vspace{-2in}
\centerline{\epsfxsize= 5in \epsfbox{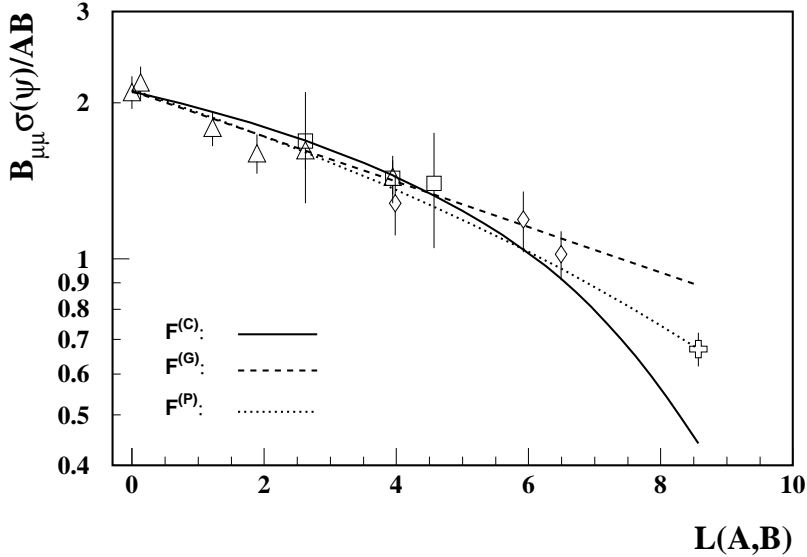}
}
\vspace{-2in}
\caption{Sensitivity of ``pre-resonance" $J/\psi$ nuclear
absorption to details  of color octet models of $c\bar{c}$
interactions in a nuclear medium from Qiu et al\protect{\cite{Qiu:1998rz}}.
The curve marked P for power law accounts for the anomalous absorption seen in
$Pb+Pb$ and deviated from Gaussian Glauber-like G expectations.}
\label{fig:qui}
\end{figure}

While the deviation from the  empirical $(AB)^{0.9}$
scaling is very clear, the dynamic origin of the effect is
far from clear in my opinion. As shown by the several curves in 
Fig.(\ref{fig:psina50}),
a $\sim$50\% drop in the $J/\psi$ yield as a function of $E_T$
is consistent
with final state co-moving hadronic absorption. While the details
are not reproduced accurately, large theoretical
uncertainties  about several key dynamical 
 ingredients preclude precise comparisons at this time. 
It is important to emphasize that the so called ``conventional hadronic''
models suffer from just as large theoretical uncertainties as
the plasma scenario models. 
Key uncertain elements include (1) the dynamical treatment of
cold nuclear absorption responsible for the ${AB}^{0.9}$ suppression
even in $p+A$, (2) the unknown hadronic $M+\psi\rightarrow D\bar{D} X$
reaction rates, and (3) the actual density evolution, $\epsilon(\vx,t)$
needed in order to make more precise calculations.
Only a schematic  ``octet model" of pre-hadronization
$c\bar{c}$ interactions in cold nuclei has been used in the above transport
models
to address the first effect.
That this is highly uncertain is shown in the work of  
Ref.\cite{Qiu:1998rz}. The observed $J/\psi$ production cross section
even without final state interactions is suppressed as follows:
\beq
\sigma_{AB\rightarrow\psi X}=\int d\sigma_{AB\rightarrow c\bar{c}}
F_{c\bar{c}\rightarrow\psi}(q^2)
\eeq{qui}
where $F$ is the formation probability of the $\psi$ from a $c\bar{c}$
that emerges from the cold nuclear target with an invariant mass
$q^2<4M_D^2$. If the pre-resonance $c\bar{c}$ 
pair multiply scatters in the nucleus
random walk would increase $q^2\rightarrow q^2 + \delta q^2 (\sigma\rho L)$
linearly with nuclear thickness as shown by the  curve G in
 Fig.(\ref{fig:qui}). This leads to an approximate
exponential suppression that can be fit well by an approximate  Glauber
nuclear absorption factor ansatz, $\exp(-\sigma_{eff}\rho L)$
if a Gaussian assumption is made.
This Gaussian model  can thus 
account for the observed $(AB)^{0.9}$ scaling light projectiles. 
{\em However},
it was shown in \cite{Qiu:1998rz} that if the power law tails due to induced
radiation in the medium is included (resulting
from the multiple 
Rutherford rescattering of the color octet $c\bar{c}$)
, then an additional nonlinear suppression
in the nuclear thickness $L$ {\em could} result (curve P). This is  because
radiation provides another way to increase the invariant mass of the
pre-resonance 
$c\bar{c}$ that can further reduce the probability 
for the pair to fit inside the $\psi$ wavefunction.
While this model is also schematic, I feel that it captures an important
element of pre-formation transient dynamical effects in nuclei
that can serve as additional
sources of nonlinear suppression, without even considering
the sought after suppression in the comoving dense matter.
The transport model curves in Fig.\ref{fig:psina50}
co-moving dissociation possesses and pre-resonance nuclear dynamics as in
Fig.\ref{fig:qui} should be combined in future  analysis
searching for the dynamical origin of
 beyond nuclear Glauber absorption in $Pb+Pb$.

Therefore, the claim of  anomalous
suppression cannot  rest therefore merely 
on generic enhanced suppression in $Pb$.
It must and has been made on the possible existence  of singular
``step-like" structure of the suppression pattern.
The evidence for "steps" is however the weakest link experimentally
because a rigorous $\chi^2$ test including the substantial systematic
errors in the $E_T$ scale has yet to be performed.
This very difficult experiment has fought valiantly for years to reduce the 
systematic errors associated with calibration of $E_T$, thick target
multiple interactions, and calibrating the Drell-Yan and other
backgrounds.
\begin{figure}
%
\centerline{\epsfxsize= 4 in \epsfbox{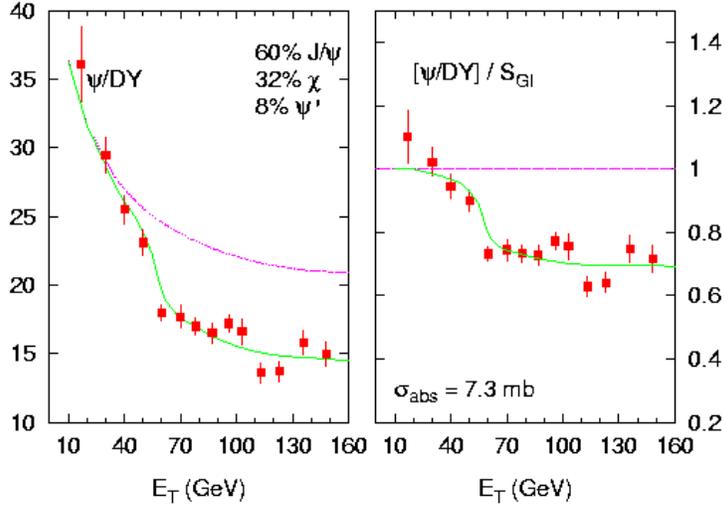}
}
\caption{Plasma scenario step function fits 
from Kharzeev et al\protect{\cite{Kharzeev:1998wa}}.
to  NA50 step-like
$J/\psi$ suppression in $Pb+Pb$ at Quark Matter 97.
Left is the $\psi$/DY ratio as a function of the $E_T$ scale at that
time. Right is the ratio normalized to the 
empirical nuclear suppression effect.}
\label{fig:dima97}
\end{figure} 
\begin{figure}
%
\centerline{
\epsfxsize= 4in \epsfbox{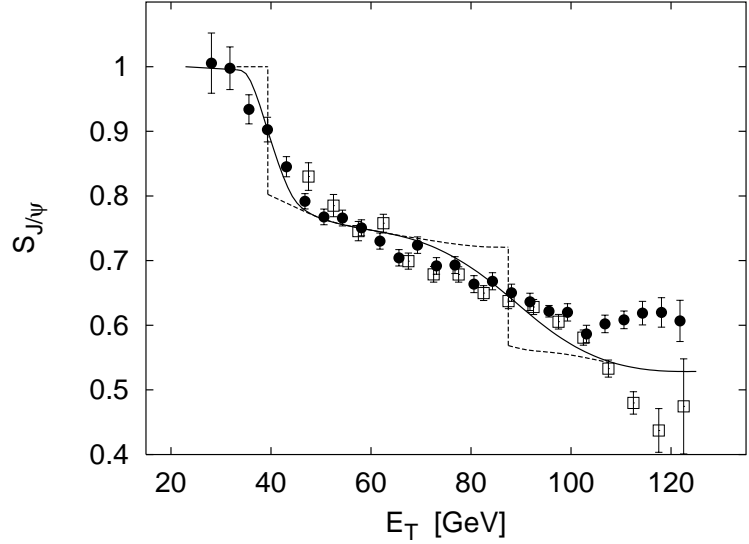}
}
\caption{Percolation model step-function fits to the $\psi$
survival propability normalized to the empirical nuclear suppression effect 
from H. Satz, Quark matter 99\protect{\cite{Satz:1999si}}.
Open square NA50 data are from thinner target runs with
$Pb+Pb$.}
\label{fig:satz99}
\end{figure} 
This is illustrated in Fig.(\ref{fig:dima97}) 
where plasma scenario fits\cite{Kharzeev:1998wa}
 to 1997 data are also shown. In Fig.(\ref{fig:satz99}) 
recent plasma scenario fits\cite{Satz:1999si}
 to the latest data show how much the data has evolved
(see \cite{Abreu:2000ni} for details). It is also clear from
Figs.(\ref{fig:dima97},\ref{fig:satz99}) that 
the plasma scenario step-function models
while capable of fitting the data cannot be
be considered as prediction of QCD. Percolation models serve
to motivate the steps.  

Nevertheless,
if the ``step-like" suppression pattern survives further experimental
scrutiny, it would certainly be the most dramatic nonlinearity observed at SPS.
In this connection, it is also important to remark
 that the energy density scale
in Fig.(\ref{fig:psina50} b) is uncertain to at least a factor
of two. There is no direct experimental
measurement of the initial energy density.
The energy scale there is infered from a RQMD model calculation.
However, as shown in Fig.(\ref{fig:ruus})  a nice hydro fit to the data
could be obtained  with an initial energy density ten times higher
than assumed in Fig.\ref{fig:psina50} b. 

Given the boldness of the NA50 claims, it is  important to  
scrutinize  both the experimental and theoretical
 foundations on which those claims are made. In science guilt is to assumed
 until  innocence is proven. 
Based on the previous discussion, I remain skeptical.

An additional problem with the plasma scenario interpretations
can be seen form the observed $E_T$ dependence of the $J/\psi$ $p_\perp$
spectra in  Fig.(\ref{fig:psipt}).
 Standard Glauber multiple collisions lead to a random walk
in transverse momentum that are expected to lead to\cite{Gavin:1988tw}
\beq
\langle p_\perp^2\rangle_{AB}= \langle p_\perp^2\rangle_{pp} + 
\frac{L}{\lambda} \delta p_\perp^2
\eeq
This is found to hold\cite{Drapier98} in all reaction including $Pb+Pb$.
In contrast, in the plasma scenario\cite{Kharzeev:1997ry}, 
only those $\psi$ are
expected  to survive that are near the surface where the nuclear depth
$L$ is small. Thus the prediction as shown in Fig(\ref{fig:psipt}) was
that the $\langle p_\perp^2\rangle$ should begin to DECREASE with increasing
$E_T$. This was not observed\cite{Drapier98,Nagle:1999ms}.
\begin{figure}
%
\centerline{\epsfxsize= 2.5 in \epsfbox{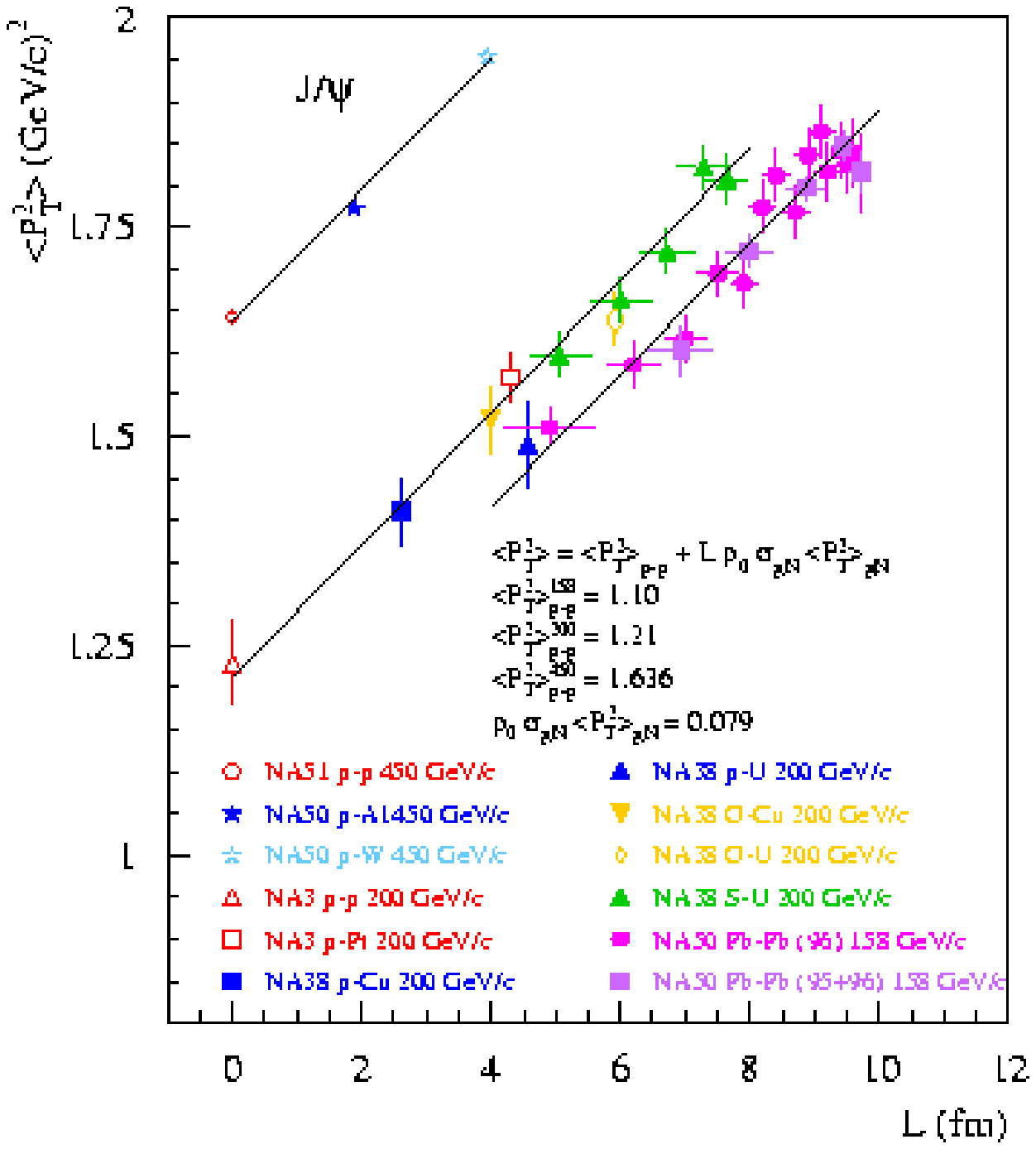}
\epsfxsize= 2.5 in 
  \epsfbox{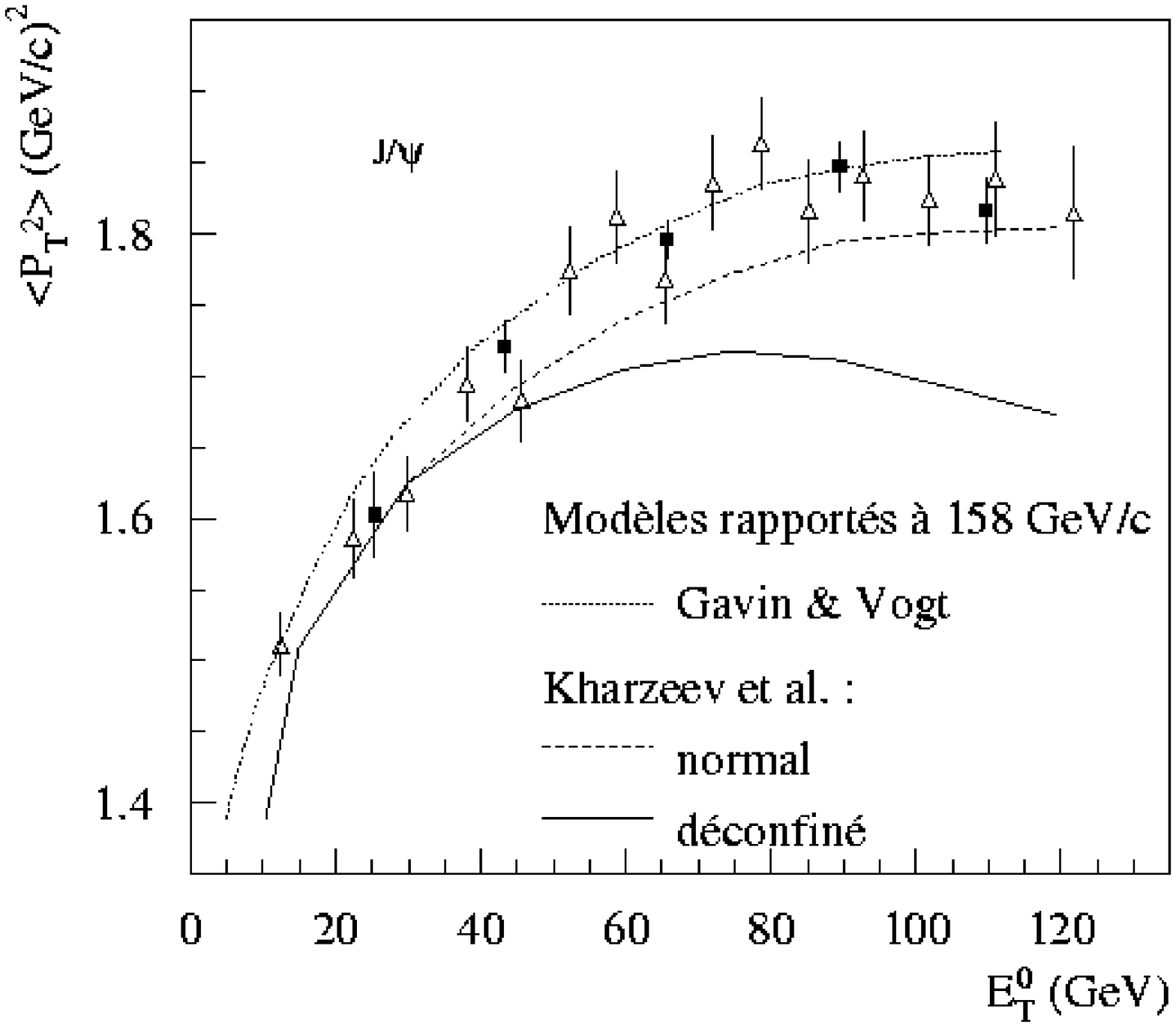}
}
\caption{The mean transverse momentum of the $J/\psi$ 
\protect{\cite{Drapier98}}
shows clear evidence of multiple scattering in the nuclear
target as expected\protect{\cite{Gavin:1988tw}} from  Glauber theory. 
Even in $Pb+Pb$ the increase of the $p_T$ is understood from the same
mechanism in contradiction\protect{\cite{Nagle:1999ms}} 
to predictions based on the plasma
scenario, where at high $E_T$,  the surviving $\psi$ are expected to have been
produced only near the nuclear surface 
regions\protect{\cite{Kharzeev:1997ry}}.
}
\label{fig:psi}
\end{figure} 
Clearly, much more work is required to sort out the very interesting
suppression pattern observed by NA38,NA50. 
Experimentally, the  claims would carry considerable more punch
if similar "step-wise" patternswere observed in other systems
,e.g. $Xe+Xe$ suitably shifted in $E_T$ 
due to the expected smaller energy densities
achieved. While it is not likely that such further
measurements will get done,  they will  be 
at RHIC. A clear prediction by H. Satz at QM99\cite{Bass:1999zq},
was  that under RHIC conditions of higher energy density,
the same step wise pattern as in Fig.(\ref{fig:satz99})
 should be observable in $Cu+Cu$
interactions. PHENIX will provide a definitive
test of this prediction soon.

\section{The High $p_\perp$ Frontier}

One of the new areas that RHIC will open in the experimental
search for the next Yukawa phase of QCD is high transverse momentum (short
wavelength) jet probes. The rates of jet production and its fragmentation
in the vacuum are well understood. The new physics here 
is the study of partonic interactions at extreme densities
through the phenomenon of jet quenching\cite{gptw}.
Final state interactions of a jet in a dense QGP are
expected to induce a large radiative energy loss\cite{mgxw}.
In fact, it was discovered in by BDMPS\cite{bdms}
 that non-Abelian energy loss
is in fact non-linear as a function of the thickness of the medium.
Tests of this and other aspects of 
 non-Abelian multiple  collision dynamical will be possible 
at RHIC\cite{glv1b}.

\begin{figure}
%
\centerline{
{\hbox{\psfig{figure=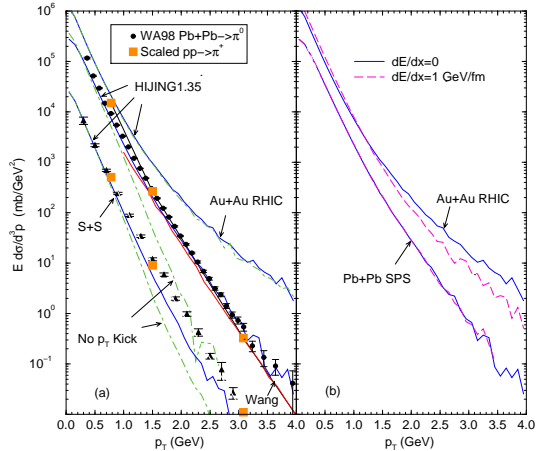,angle=-90,width=7cm}}}
}
\caption{Jet quenching at SPS 
vs RHIC from ref.\protect{\cite{Gyulassy:1998nc}} compared to WA98 data.
At RHIC the power law tail extends finally far enough above 
 the nonperturbative
``noise" to make jet quenching  observable.
}
\label{fig:jets}
\end{figure}

At SPS energies, this  physics is out of reach given the
dominance of  nonperturbative effects as shown in Fig.(\ref{fig:jets})
from ref. \cite{Gyulassy:1998nc}. HIJING accidently fits the WA98
data with or without jet quenching. At the SPS no clean separation of
soft and hard dynamics is kinematically possible.
However, at RHIC energies, the power law tails of the single inclusive
distributions stick far enough above the confusing soft "noise" to gain
sensitivity to the form of the non-Abelian $dE/dx$ as seen
 in Fig.(\ref{fig:jets}).
This problem is closely related also to the problem
of pre-resonance $J/\psi$ absorption discussed previously\cite{Qiu:1998rz}.
Understanding jet quenching is also prerequisite in testing
the dynamical assumptions of recent  covariant
parton transport theories\cite{molnar,Gyulassy:1997zn}.

\section{Summary}

The next Yukawa phase is awaiting discovery. The SPS data have
provided many intriguing indirect hints that new physics
is operating in dense matter. Many puzzles, claims and counter claims
remain because at SPS  energies, both hadronic and partonic
models have partial overlapping domains of validity.
This ``duality" is analogous to the problem of interpreting
the R factor in $e^+e^-$ collisions
below $\surd s<10$ GeV. The ratio of hadronic to leptonic
production cross sections only reaches the the magic 11/3 of pQCD
above that threshold region where the $s\bar{s},c\bar{c},b\bar{b}T$
vector mesons dominate the nonperturbative hadronic physics.
Similarly SPS is at the door-step where hardonic resonances begin to melt away
and a pQCD continuum description start to become more relevant.

With RHIC the factor of ten increase in the initial energy density will be
unambigously in the QGP continuum. The matter so formed will also have much
more time to develop collective signatures.  The factor of ten smaller
wavelength probes will finally allow experimentalists to resolve (i.e. see) the
quark and gluon degrees of freedom of that plasma. Direct observation of
longitudinal work, transverse azimuthal collectivity, time delay, step-wise
$J/\psi$ suppression in $Cu+Cu$, and jet quenching will offer direct signatures
of the sought after new phase of QCD matter.

\section*{Acknowledgements}
I thank  Profs. H. Horiuchi and T. Hatsuda 
for organizing this Yukawa symposium.
I would especially acknowledge 
Mayor Baba and the city of Nishinomia their support
of this symposium series.
This work was supported by the Director, Office of Energy Research,
Division of Nuclear Physics of the Office of High Energy and Nuclear Physics
of the U.S. Department of Energy under Contract No. DE-FG-02-93ER-40764.

\end{document}